\documentclass[prx, twocolumn, showpacs, preprintnumbers, amsmath, amssymb, superscriptaddress, aps,longbibliography]{revtex4-2}
\usepackage{amssymb}
\usepackage{latexsym}
\usepackage{graphicx}
\usepackage{wasysym}
\usepackage{relsize}
\usepackage{hyperref}
\usepackage[english]{babel}
\hypersetup{linktocpage,,colorlinks=true}
\usepackage[dvipsnames]{xcolor}
\usepackage{color}
\usepackage{epsfig}
\usepackage[utf8]{inputenc}
\usepackage[T1]{fontenc}
\usepackage{newtxtext,newtxmath}
\usepackage{pdfrender}
\colorlet{RED}{red}

\newcommand*{\boldgreek}[1]{%
  \textpdfrender{%
    TextRenderingMode=FillStroke,%
    LineWidth=.35pt,%
  }{#1}%
}
\usepackage{dcolumn}
\usepackage{amsmath}
\usepackage{amsfonts}
\usepackage{bm}
\usepackage{mathtools}

\newcommand{\dg}{^{\dagger}}
\newcommand{\be}{\begin{equation}}
\newcommand{\ee}{\end{equation}}
\newcommand{\bea}{\begin{eqnarray}}
\newcommand{\eea}{\end{eqnarray}}

\newcommand{\s}{\sigma}
\renewcommand{\S}{{\cal P}}

\newcommand{\red}[1]{{\color{red}#1}}
\newcommand{\bk}{{\bf k}}
\newcommand{\bR}{{\bf R}}
\newcommand{\br}{{\bf r}}
\newcommand{\bx}{{\bf x}}
\newcommand{\by}{{\bf y}}
\newcommand{\pmat}[1]{\begin{pmatrix}#1\end{pmatrix}}
\newcommand{\dw}{\downarrow}
\newcommand{\up}{\uparrow}
\newcommand{\vu}{v (\bx )}
\newcommand{\ltappr}{{{\lower4pt\hbox{$<$} } \atop \widetilde{ \ \ \ }}}

\newlength{\figwidth}
\newlength{\figfull}
\newcommand{\fg}[3]
{\begin{figure}[tb]\vspace*{-0cm}\centerline{\includegraphics[width=\figwidth]{#1}}\vskip
-0.2cm \caption{#3}\label{#2}\end{figure}}
\newcommand{\fgwide}[3]
{\onecolumngrid

\begin{figure}[h]\vspace*{-0cm}\centerline{\includegraphics[width=\figfull]{#1}}\vskip
-0.2cm \caption{#3}\label{#2}\end{figure}\twocolumngrid
}
\newcommand{\fgh}[3]
{\begin{figure}[hbt]\vspace*{-0cm}\centerline{\includegraphics[width=\figwidth]{#1}}\vskip
-0.2cm \caption{#3}\label{#2}\end{figure}}

\newcommand{\fgb}[3]
{\begin{figure}[b]\vspace*{-0cm}\centerline{\includegraphics[width=\figwidth]{#1}}\vskip
-0.2cm \caption{#3}\label{#2}\end{figure}}
\begin{document}
\title{Order Fractionalization  in a  Kitaev-Kondo model }
\author{ Alexei M. Tsvelik}
\affiliation{Division of Condensed Matter Physics and Materials Science, Brookhaven National Laboratory, Upton, NY 11973-5000, USA}
\author{Piers Coleman}
\affiliation{Department of Physics and Astronomy, Rutgers University,
Piscataway, NJ 08854, USA}
\affiliation{Department of Physics, Royal Holloway University of
London, Egham, Surrey TW20 0EX, UK}
 \date{\today } 
 
 \begin{abstract} 
 We describe a mechanism for order fractionalization in a
two-dimensional Kondo lattice model, in which electrons interact with
a gapless spin liquid of Majorana fermions described by the
Yao-Lee (YL) model.  
When the Kondo coupling to the conduction electrons exceeds a
critical value,  the model develops a superconducting
instability into a state with a a spinor order parameter with 
charge $e$ and spin $S=1/2$. The broken symmetry state develops a
gapless Majorana Dirac cone in the bulk.
By including an appropriate gauge string, we can 
show that the charge $e$, spinorial order develops 
off-diagonal long range order that allows electrons to coherently tunnel
arbitrarily long distances through the spin liquid. 
\end{abstract}


\maketitle
%
\section{Introduction}\label{}
%
%
   One of the fascinating properties of quantum materials is the
phenomenon of fractionalization, whereby excitations break-up into
emergent particles with fractional quantum numbers.
Well-established examples of fractionalization include 
anyons in the quantum Hall effect and the break-up of 
magnons into $S=1/2$ spinons
in the one dimensional Heisenberg spin chain.  There is great current
interest in the possibility that new patterns of fractionalization can
lead to new kinds of quantum phases and quantum materials.

\figwidth=0.5\textwidth
\fgh{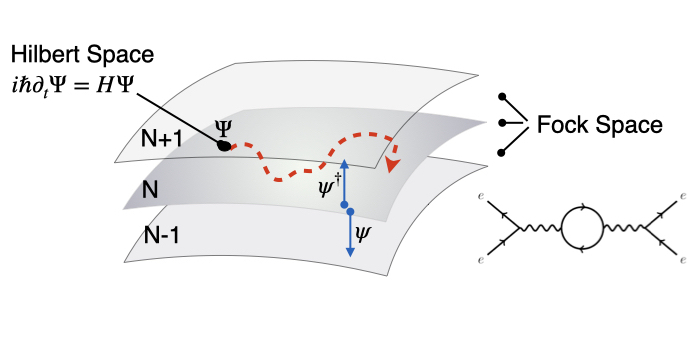}{ladders}{In second quantization, the
``physical'' Hilbert space of definite particle numbers is expanded to
a  Fock space that allows the description of particle fields.}
There are some important
parallels between second quantization and fractionalization. 
We recall that 
even though a many-body electron wave function
evolves in a Hilbert space of rigorously 
fixed particle number, physical quantities such as density
\begin{eqnarray}\label{l}
\rho (\bx ) = \sum_{j}
\delta (\bx  - \bx_{j}) 
\longrightarrow \psi \dg (\bx )\psi (\bx ),
\end{eqnarray}
factorize into creation and annihilation operators $\psi\dg (x)$
and $\psi (x)$ that link Hilbert spaces of different particle number (see
Fig. \ref{ladders}). 
Thus the description of particles requires
an expansion of the Hilbert space
into a larger Fock space. Normally we take this for granted - 
we are  quite accustomed to the notion that photons create
particle-hole pairs, content in the understanding that gauge
invariance ($\psi\rightarrow e^{i\alpha (x)}\psi $, $A\rightarrow
A+\nabla \alpha (x)$) 
preserves particle number. 
\fg{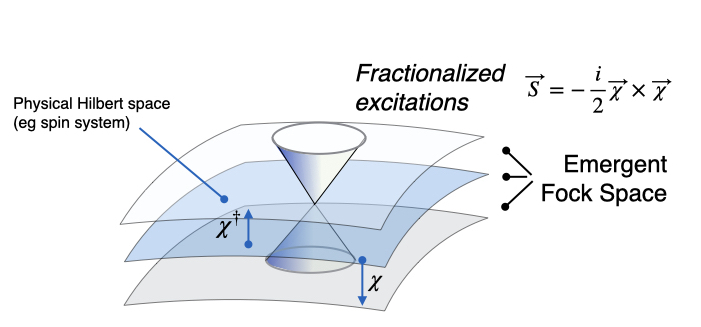}{emergent}{Fractionalization involves the
break-up of physical operators such as spin, into  excitations with
fractional quantum numbers which require an emergent Fock space for
their description.}
In a similar fashion, fractionalization can be regarded as 
an {\sl emergent  second-quantization}, in which
the microscopic variables, such as the spin, 
factorize into operators that describe fractionalized 
quasiparticles (see Fig. \ref{emergent}), thus in a spin liquid 
a spin flip creates a pair of spinons. 
Such fractionalized particles live within an emergent Fock space, 
and like their vacuum counterparts, move under the influence of 
a gauge field which preserves the constraints of the physical Hilbert space.

One of the most dramatic manifestations of second-quantization
is the formation of superfluid  condensates, in which the field operators
develop off-diagonal long-range order (ODLRO), 
manifested as a factorization of the density matrix in terms of the
order parameter $\langle \psi (\bx)\rangle =\Psi (\bx)$,
\begin{equation}\label{}
\langle \psi (\bx ) \psi \dg (\by)
\rangle \xrightarrow[]{|\bx-\by|\rightarrow \infty} 
 \Psi(\bx)\Psi (\by)^{*}.
\end{equation}
The description of superconductors is more subtle, 
for now the condensate field operator carries charge and  
transforms under a gauge transformation $\hat \psi (x)\rightarrow e^{i\alpha
(x) }\hat \psi (x)$,
$\vec{ A}\rightarrow \vec{ A}+ 
\nabla \alpha $, so that 
$\langle \Psi (x)\rangle$ vanishes after averaging over the
gauge fields, a result known as Elitzur's theorem.
A sharper definition of ODLRO\cite{Hansson_2004}
then requires 
the introduction of a gauge
invariant boson field\cite{Dirac1955} 
\begin{eqnarray}\label{l}
\hat \psi_{N} (\bx) = \hat \psi (\bx) 
e^{-i \int d^{3}r \vec A (\br )\cdot\vec{
E}_{cl} (\br-\bx)
},
\end{eqnarray}
where  $\vec{E}_{cl} ({\bf r}) $ is the
classical electric field of a point charge at the origin,
i.e $\nabla \cdot E_{cl}= \delta ({\bf r} )$, $\vec{E}_{cl} (\br) = \hat \br/
(4 \pi r^{2})$), and $\vec{A}$ is the fluctuating, quantum vector
potential.
Off-diagonal 
long-range order is then defined by 
\begin{eqnarray}\label{l}
\langle \hat \psi_{N}(\bx  )
\hat \psi_{N}\dg
(\by)\rangle 
 \xrightarrow{|\bx-\by|\rightarrow\infty} \Psi (\bx )
\Psi (\by )^{*},
\end{eqnarray}
where $\Psi (\bx )= \langle \hat  \psi_{N} (\bx )\rangle $.
The massive nature of the vector potential inside a Meissner phase,
guarantees that this result holds true, even when quantum fluctuations
are included. 
{ These considerations lead us to ask: if fractionalization is a kind of emergent second-quantization,
is there a fractionalized analog of superfluidity or
superconductivity?

Early theoretical studies of a possible
interplay between fractionalization and
broken symmetry were inspired by 
the RVB-theory of cuprate
superconductivity\cite{baskaran,ioffelarkin,ioffekotliar}; 
in these papers fractionalization 
appears under the guise of ``spin-charge
separation''. In particular,  the appearance of a charge $e$ boson
in the slave-boson decomposition of the
electron operator $c\dg_{j\sigma }\rightarrow f\dg_{j\sigma }b_{j}$
raised the early intriguing possibility of novel order parameters associated
with spin-charge separation. 
Although the presence of an emergent $U (1)$
gauge field associated with spin-charge separation 
appeared to forstall a superconducting, charge $e$ condensate, it was
soon realized\cite{wenzee,sachdev92} that there might be a topological effect. 
In particular, the topological interplay of the electromagnetic and emergent
$U (1)$ gauge fields, Wen and Lee\cite{wenzee}
identified a two-parameter family of
vortices and subsequently, Sachdev proposed a possible stablization of $h/e$
vortices\cite{sachdev92} near a superconductor, pseudo-gap phase
boundary. The modern term 
``fractionalization'' appeared in a second-generation of
theories\cite{balentsfishernayak,senthilfisher} that were inspired by the
pseudo-gap phase of cuprate superconductors.  These theories 
identified the fractionalization of electrons and spins with an emergent 
$Z_{2}$ gauge field. Senthil and Fisher\cite{senthilfisher} introduced
the term ``vison'' to describe the vortices of the $Z_{2}$ field. 
In their theory, the development of a gap in the vison spectrum 
gives rise to a novel fractionalized insulator gapped vison
excitations. }

{Our paper returns to these early lines of
investigation,  taking crucial
advantage of the Kitaev approach to introduce a new platform for the
discussion, in the form of a family of models 
which control the gauge fields that are 
at the heart of fractionalization. The
Kitaev approach with its static $Z_{2}$ gauge fields 
now makes it explicitly clear that fractionalization is 
physical.  
This then leads us 
to reconsider the question of 
whether the condensation of fractionalized bosons 
can actually give rise to novel forms of order parameter?}
This could happen, for
instance if a spinon binds to an electron. 
The resulting order parameter has the potential 
to carry fractional quantum numbers with novel order-parameter
topologies and symmetries,  giving rise to  a conjectured ``order
fractionalization''\cite{komijani2019order}.

Here,  we explore the idea of order fractionalization
within the context of the Kondo lattice model. 
The Kondo lattice has a venerable history: first written down by
Kasuya in 1955\cite{kasuya1955},  later proposed in the 1970s by Mott
and Doniach\cite{Mott:1974ui,doniach} to explain heavy fermion
materials. The Kondo lattice describes a 
lattice of local moments, coupled to 
conduction  electrons via an antiferromagnetic 
super-exchange of strength $J$.
When $J$ is sufficiently large, the local moments become screened 
by conduction electrons, 
liberating their  entangled spin degrees of freedom into the conduction sea as
a narrow band of ``heavy electrons''.

From a modern perspective, 
the  Kondo lattice effect can be understood as a spin
fractionalization of localized moments. In a heavy Fermi liquid, 
local moments
split into spin 1/2 heavy
fermions, conventionally described as a 
bilinear of S=1/2  ``Dirac'' fermions\cite{read1983,Coleman1983, auerbach,colemanandrei,senthil2003},
\begin{equation}\label{}
\vec{S} (\bx_{j}) \rightarrow  f\dg_{j\alpha }\left(\frac{\vec{\sigma}}{2} \right)_{\alpha \beta }f_{j\beta }.
\end{equation}
In this scenario, a spin flip creates 
a pair of ``spinons'' moving in an emergent $U
(1)$ gauge field
 which enforces their incompressibility\cite{colemanprl1987}. 
When the Kondo effect takes
place, the coherent exchange of spin between the electron and spin
fluid Higgses the $U (1)$ gauge field, 
locking it to the 
electromagnetic field and converting the neutral spinons into 
charged heavy fermions\cite{coleman2005}.
This ``Dirac fractionalization'' of spins
provides a natural way to understand the
expansion of the Fermi surface in the Kondo lattice, described by Oshikawa's theorem\cite{oshikawa2000,hazra2021}.
\fg{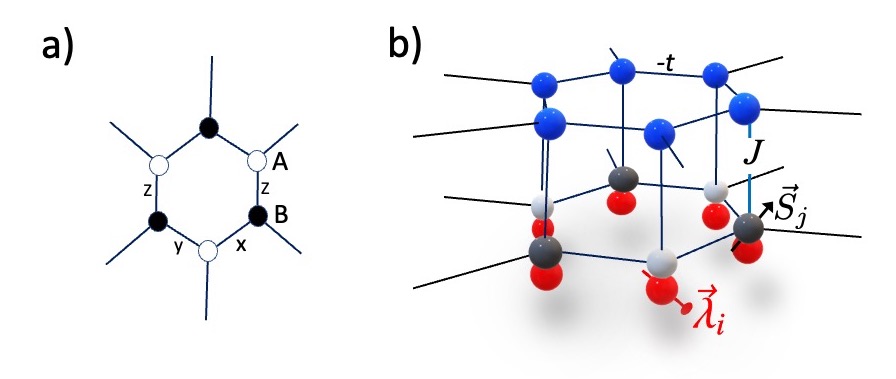}{fig1schem}{Schematic of the Kitaev Kondo
model showing a), the x, y and z bonds of the lattice and b)
the structure. The 
lower layer is a Yao-Lee  spin liquid with gapless spin
($\vec{S}_{j} $) and gapped orbital degrees of freedom ($\vec{\lambda }_{i}$). The upper layer is a honeycomb lattice of conduction
electrons, coupled to the spin liquid via a Kondo interaction. }

Here we study an  unconventional spin fractionalization 
into Majorana fermions, first proposed in\cite{coleman94,Coleman_PhysicaB1993}
\begin{equation}\label{}
\vec{S} (\bx_{j}) \rightarrow - \frac{i}{2} (\vec{ \chi}_{j}\times \vec{ \chi }_{j}),
\end{equation}
where $\vec{\chi }_{j}=
 (\chi^{1}_{j},\chi^{2}_{j},\chi^{3}_{j})$ is a spin-1 majorana\footnote{Note that throughout this paper, following
contemporary physics usage, we shall use the term
{\sl majorana} as a proper noun to describe a Majorana fermion, thus 
using a lower-case ``m''.}
 that
 moves in a $Z_{2}$ gauge field. 
In this alternate scenario, a spin flip 
produces a pair of majoranas. 
Majorana fractionalization gives rise to a 
gapless band of neutral excitations, and  has been proposed as a driver of 
odd-frequency pairing\cite{coleman94} and the origin 
of Kondo insulators with neutral Fermi
surfaces\cite{Coleman_PhysicaB1993,Baskaran_arxiv2015,Erten:2017bj}.

A novelty of our work, is the Kondo-coupling of electrons to a gapless
spin liquid 
in which Majorana fractionalization is
rigorously established. 
We combine a variant of the Kitaev honeycomb model\cite{Kitaev_honeycomb}, called the
Yao-Lee  model\cite{yaolee11,kivelson09,arovas09,seifert2018,janssen20,janssen21}, with a
corresponding lattice of mobile electrons. 
{ Like the Kitaev model, 
the Yao-Lee model is exactly solvable,
 which allows a nonperturbative treatment of the fractionalization,
 i.e the strongest correlations in the model. 
The weaker Kondo exchange is then treated withing mean field
approximation in the manner of
Bogoluibov-de-Gennes theory. } 
Unlike the original Kitaev model, in which spin excitations
create gapped $Z_{2}$ vortices\cite{Kitaev_honeycomb}, 
the Yao-Lee model describes a spin liquid 
in which spin flips fractionalize into gapless  
Majorana fermions, leaving the static  $Z_{2}$ gauge field unaffected. { This  radically affects  the character  of the Kondo interaction between the 
conduction electrons and the local moments, opening up the possibility
of a fractionalized order parameter 
formed from a pair condensation of electrons and Majorana fermions.}  

In the Yao Lee model, the motion of the Majorana fermions is described by the Hamiltonian
\begin{equation}\label{ksl1}
H_{YL}
=K\sum_{<i,j> 
} u_{ij}
(i \vec\chi_i\cdot \vec\chi_j),
\end{equation}
where $u_{ij}=\pm 1$ is the static gauge field. 
The exchange-coupling of a Yao-Lee spin liquid 
to electrons on an adjacent honeycomb layer (Fig. \ref{fig1schem})
now forms a Kondo lattice
in which the absence of gauge fluctuations establishes 
an order-fractionalized 
state\cite{komijani2019order} in which 
electrons and majoranas combine into 
charge $e$, $S=1/2$ bosons
\begin{equation}\label{vop}
\hat v (\bx_{j}) =
 \bigl(\vec \sigma_{\alpha \beta } \cdot \vec \chi_j\bigr)c_{j\beta}= \begin{pmatrix}
\hat v_{j\uparrow}\\ \hat v_{j\downarrow} \end{pmatrix},
\end{equation}
where 
$c_{j}$ is an electron operator 
at site $j$.
When this boson condenses, it gives rise to a state
in which triplet pairs have fractionalized into condensed 
bosonic spinors, forming
a well-defined order parameter with charge e and spin 1/2.

Since the fractionalized fields $\vec{\chi}_{j} $ and $\hat v (x)$ 
carry a $Z_{2}$ charge, 
a gauge invariant definition
of off-diagonal long range order follows a similar procedure to a
superconductor, introducing a 
string of $Z_{2}$ gauge fields, 
\begin{equation}\label{Z2string}
\S (\bx_{i},\by _{j}) = \prod_{l\in P_{{j\rightarrow i}}}u_{(l+1,l)}
\end{equation}
along a path $P_{j\rightarrow i}$ linking sites $j$ and $i$\footnote{Here we have introduced a bracket around the indices $l+1$
and $l$ to imply that the two indices are always re-ordered so that the
A sublattice comes first.}, giving
rise to the  
asymptotic factorization
\begin{equation}\label{}
\langle  v ({\bx})\S (\bx,\by) v\dg  ({\by}) 
\rangle \xrightarrow[]{|\bx-\by|\rightarrow \infty} 
v (\bx) v \dg (\by).
\end{equation}
\fgwide{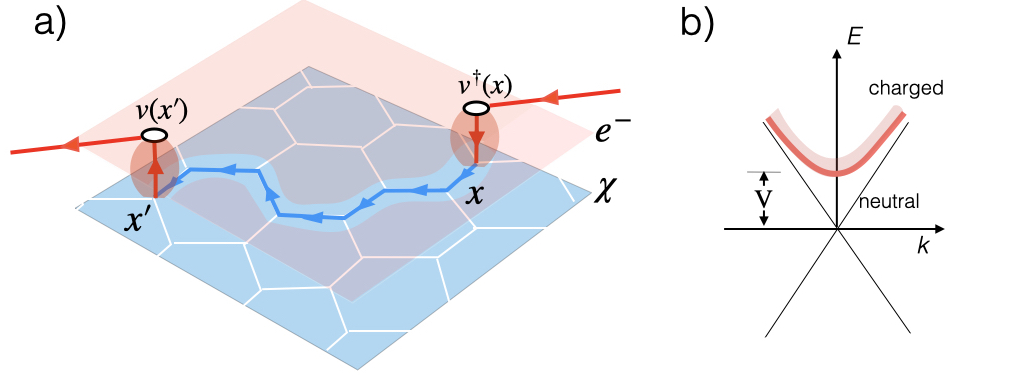}{figtunnel}{a) Development of a charge $e$
condensate $v(x)$ permits the coherent tunneling of electrons
through the spin-liquid over arbitrary distances. b) The mismatch
between the three Majorana components of the spin liquid and the four Majorana components
of the conduction electrons \eqref{split} leads to a decoupled ``neutral''
conduction electron Majorana Dirac cone with a gap to charged electron
excitations. }
\noindent As in a superconductor, the $Z_{2}$
vortices, or ``visons'' are 
absent in the ground state, so one can adopt an
axial gauge where the $u_{(i,j)}=1$ and the string becomes unity. 
The development of this off-diagonal long
range order with fractional quantum numbers constitutes order
fractionalization\cite{komijani2019order,wugalter20}.

One of the most
dramatic consequences of this off-diagonal long-range order, is that
electrons can tunnel 
over
arbitrarily large distances through the spin liquid (Fig. \ref{figtunnel}a). The
amplitude $\Sigma (x',x)$ for this process is
directly proportional to the spinor
order at the entry and exit points $x$ and $x'$,
\begin{equation}\label{}
\Sigma (x',x) \sim v(x') \frac{1}{|x'-x|^{2}}v\dg  (x).
\end{equation}
The long-range coherence of this process reflects the
order-fractionalization.
The neutral character of the Majoranas in the spin fluid has two
interesting consequences: first, it means that when electrons emerge
from this tunneling process, they can reappear into the conduction
fluid as either electrons, or holes, giving rise to both normal and
Andreev scattering processes. Secondly, the mismatch between the quantum numbers of the electrons and the
Majorana fermions then 
gaps out three Majorana
components of the conduction sea, leaving behind 
a neutral Majorana cone of conduction-like excitations (Fig. \ref{figtunnel}b).   The sharp coherence
of this neutral band reflects the phase coherence of the charge $e$
spinor order.


If we sample the spinor field locally, we can construct 
a composite order parameter
\begin{eqnarray}\label{triad}
\langle \hat v^{T} (\bx) i \sigma_{2}\vec{\sigma }v (\bx )\rangle  &\propto&
\langle c_{\uparrow}(\bx)c_{\downarrow}(\bx) \vec{ S} (\bx  )\rangle\cr
&=&{\Psi}(\bx)(\hat d_{1} (\bx)+ i \hat d_{2} (\bx)),
\end{eqnarray}
representing the local binding of a
Cooper pair to a spin
where $\hat d^{1}$ and $\hat  d^{2}$ are members of an orthogonal
triad of unit vectors $ (\hat d^{1},\hat  d^{2}, \hat d^{3})$.
However, this is not the primary order parameter of the physics, as
can be seen by observing that the gap in the spectrum is proportional
to $|v|$, rather than $|v|^{2}$. 

The structure of this paper is as follows.  Section \ref{sec2} introduces the
Kitaev Kondo Model.  Section \ref{sec3} reviews the properties of the Yao-Lee
spin liquid. Section \ref{sec4} introduces the mean-field theory for the
order-fractionalizing transition.  Section \ref{sec5} discusses the quantum
critical transition at half filling and the first order order
fractionalizing transition that develops with finite doping.  Section
 \ref{sec6} discusses the nature of the off-diagonal fractionalized long-range
order. Section  \ref{sec7} discusses the nature of the triplet pairing, and its
odd-frequency character.  Section  \ref{sec8} discusses the phase diagram and
long-wavelength action. Finally, Section  \ref{sec9} discusses the broader
implications of our results. 


%
%
%

%
%
%
%
%
%
%
%
%
%
%
\section{Kitaev-Kondo model}\label{sec2}

Coupling a Yao-Lee spin liquid to a conduction sea,  
forms a Kondo lattice
with Hamiltonian 
$H = H_{C}+H_{YL}+H_{K}$  (see Fig. \ref{fig1schem}), where 
\begin{eqnarray}\label{l}
H_{C}&=& -t\sum_{<i,j>} ( c\dg_{i\sigma }c_{j\sigma }+{\rm H.c}) -\mu \sum_{j}c\dg_{j\sigma }c_{j\sigma },\label{H0}\\
H_{YL} &=&(K/2)
\sum_{<i,j>}
(\vec \s_i\cdot \vec \s_j)\lambda_i^{\alpha_{ij}}\lambda^{\alpha_{ij}}_j,\label{SL}\\
H_{K} &=& J \sum_{j} \vec{S}_{j}\cdot c\dg _{j}\vec{ \sigma }c_{j}.\label{HK}
\end{eqnarray}
Here $\langle i,j\rangle $ denotes a pair of neighboring sites, with
$i$ on the even (A) and $j$ on the odd (B) sublattice. 
$H_{C}$ is a tight-binding model of conduction electrons moving on
honeycomb lattice with hopping matrix element $-t$.
$H_{YL}$ is the Yao-Lee model, a version of
the Kitaev honeycomb model in which each site has both an orbital
degree of freedom, denoted by three Pauli orbital 
$\lambda^{a}_{j}$($a=1,2,3$) operators and 
a spin 
degree of freedom, denoted by the Pauli matrices  $\vec \sigma _{j}$.
\footnote{The antiferromagnetic sign of the spin-orbital coupling $+K$ differs from the canonical choice in Kitaev
honeycomb models, is chosen to simplify string operators in later work.}
The ${\vec{S}}_{j} \equiv \vec{\s}_{j}/2 $ are the normalized
spins for the localized moments and the $\alpha_{ij}=x,y,z$ along the x,y
and z bonds of the honeycomb lattice (Fig. \ref{fig1schem}).  Finally,
$H_{K}$ describes an antiferromagnetic exchange interaction
between the electrons and the spin liquid.  

Several earlier variants of Kitaev Kondo lattices
have been 
considered, including models that couple the original, spin-gapped
Kitaev spin liquid to a conduction sea\cite{seifert2018,choi2018,Freire21}, and models
that couple a Yao-Lee spin liquid to a conduction sea via an
anisotropic, octupolar coupling\cite{miranda20}. 
The current model builds on these earlier treatments, isotropically 
coupling electrons to a solvable gapless spin-liquid to 
preserve  the $SU (2)$ spin symmetry, leading to a fluid in which
crucially, the gapped visons decouple from the low energy spin and charge fluctuations.

\section{The Yao-Lee Spin Liquid}\label{sec3}
%
We begin by recapitulating the key features of the Yao-Lee spin liquid\cite{yaolee11,kivelson09,arovas09,seifert2018,janssen20,janssen21}.
The first step is to transmute the spin and orbital operators into
fermions, 
expanding the
Hilbert space into a Fock space by adding two
ancilliary Majorana fermions  $\Phi^{S}_{j}$ and $\Phi^{T}_{j}$,
equivalent to one ancilliary qubit, at each 
site.
We use the normalizing convention 
$(\Phi^{S,T}_{j})^{2}=\frac{1}{2}$ throughout this paper for all majoranas.
The spin and orbital 
majoranas
 are defined as a fusion of the Pauli operators with the
ancillars\cite{CMT95} 
\begin{equation}\label{defined}
\chi_{j}^{\alpha } = \Phi^{S}_{j}\s
_{j}^{\alpha },\qquad  b^{a}=\Phi^{T}_{j}\lambda^{a}_{j}. 
\end{equation}
These satisfy canonical anticommutation algebras
$\{\chi^{a}_{j},\chi^{b}_{j} \}=\{b^{a}_{j},b^{b}_{j} \}
= \delta^{ab}\delta_{ij}$ and $\{\chi^{a}_{j},b^{b}_{k} \}= 0$.
Using the fact that
$\sigma^{x}\sigma^{y}\sigma^{z}=\lambda^{x}\lambda^{y}\lambda^{z}=i$, we obtain
the reverse transformations 
\begin{eqnarray}\label{l}
\Phi_{j}^{S}=-2 i \chi_{j}^{1}\chi_{j}^{2}\chi_{j}^{3},\qquad 
\Phi_{j}^{T}=-2 i b_{j}^{1}b_{j}^{2}b_{j}^{3},
\end{eqnarray}
which enable us to write the spins and orbitals as 
\begin{eqnarray}\label{key}
\vec{\s}_{j}&=&
2\Phi_{j}^{S}\vec{\chi}_{j}=-{i}\vec{\chi}_{j}\times
\vec{\chi}_{j},\cr
\vec{\lambda }_{j}&=&
2\Phi_{j}^{T}\vec{b}_{j}=-{i}\vec{b}_{j}\times
\vec{b}_{j},
\end{eqnarray}
where we use vector notation $\vec{\chi }_{j}= (\chi^{1}_{j},\chi^{2}_{j},
\chi^{3}_{j})$
and $\vec{b }_{j}= (b^{1}_{j},b^{2}_{j},
b^{3}_{j})$. It follows that 
$\s^{a}_j\lambda_{j}^{\alpha }= - 2i \hat D_{j}
\chi^{a}_{j}b_{j}^{\alpha }
$,
where 
\begin{equation}\label{constraint}
\hat D_{j}=-2 i \Phi_{j}^{S}\Phi_{j}^{T}=8i
\chi_{j}^{1}\chi_{j}^{2}\chi_{j}^{3}b_{j}^{1}b_{j}^{2}b_{j}^{3}.
\end{equation}
Now $\hat D_{j}$, with eigenvalues $D_{j}=\pm 1$,
commutes with $H$, and the constraint
$D_{j}=1$ 
selects a physical Hilbert space 
\begin{equation}\label{}
\vert \psi_{p}\rangle = \prod_{j}\frac{1}{2} (1+D_{j})\vert \psi \rangle .
\end{equation}
in which 
\begin{equation}\label{}
\s_{j}^{a}\lambda_{j}^{\alpha }= -2i \chi_{j}^{a}b^{\alpha }_{j},
\end{equation}
which enables us to rewrite \eqref{SL} in the expanded Fock space of
majoranas as
\begin{equation}\label{ksl}
H_{YL}
=K\sum_{<i,j> 
}\hat u_{ij}
(i \vec\chi_i\cdot \vec\chi_j).
\end{equation}
Here, the $i$ and $j$ sites are on the A and B sublattices
respectively, and
the 
$ \hat  u_{i,j}=-2i 
b_{i}^{\alpha_{ij}}b_{j}^{\alpha_{ij}}$
are gauge fields that live on
the bonds, with eigenvalues $u_{ij}=\pm 1$. {(Note the negative sign
in the bond operator, which simplifies later
calculation of string variables).}
The remarkable 
feature of Kitaev models, making them so useful for our analysis, 
is that the gauge fields $\hat u_{ij}$ are completely static
variables, rigorously commuting with $H$ and the constraint operators $D_{j}$.

Like the Kitaev honeycomb model, $H_{YL}$ describes a $Z_{2}$ spin
liquid with the gauge symmetry
\begin{equation}\label{}
\vec{\chi }_{j}\rightarrow Z_{j}\vec{\chi }_{j}, \quad \hat u_{ij}\rightarrow
Z_{i}\hat u_{ij}Z_{j}, \qquad (Z_{j}=\pm 1).
\end{equation}
The product
\begin{equation}\label{plaquet}
\hat W_{\rm p}= \prod_{<l+1,l>\in\rm p} \hat u_{(l+1,l)}= \prod_{j\in\rm p} \lambda^{a_{j}}_{j}
\end{equation}
around a hexagonal plaquette $p$, where the $a_{j}$ denote the directions
exterior to the plaquette,  commutes with $H$ and the $D_{j}$, forming 
a set of gauge-invariant constants describing static $Z_{2}$ fluxes
(visons).  Note our use of the parentheses around the indices of the
$\hat  u_{(l+1,l)}$, 
which re-arranges
the subscripts so that the A sublattice index is first. 

In the ground state,  the eigenvalues $W_{p}=1$ on every plaquette while
plaquettes where $W_{p}=-1$ describe gapped ``vison''
excitations. 
Since a $\vec\lambda_{j}$ flips a bond, it creates
two visons, so the orbital degrees of freedom are gapped. This is then
a {\sl Higgs phase} for the $Z_{2}$ fields, in which the $Z_{2}$ gauge
field has become massive. 
However, unlike the Kitaev honeycomb model,  there are three
gapless 
$\vec{\chi }$ majoranas which describe the fractionalization of the spins
$\vec{S}_{j}= -(i/2)\vec \chi_{j}\times \vec{ \chi}_{j}$. 
The action of $\vec{ S}_{j}$ does not create visons, leading to a 
spin liquid with gapless spin excitations.

Although the vector Majorana fermions are not
gauge invariant, they can be made so by 
attaching a gauge string to them to create a $Z_{2}$-neutral field
\begin{equation}\label{}
\vec{\chi }_{N} (i) = \vec{\chi} (i)\S (\bx_{i},-\infty )
\end{equation}
where $\S (\bx ,\by  )
$ is the $Z_{2}$ string defined in \eqref{Z2string} . 

 It is convenient to choose a axial gauge in which the
$u^{x,y}_{ij}=1$ along the x and y bonds, leaving the z-bonds as the 
dependent variables. \footnote{Note that to avoid the situation where
we have a row of $u_{ij}=-1$, leading to an ambiguity in the
definition of the $\chi_{N}$, we choose periodic boundary conditions
where the horizontal x and y bonds wrap around the torus to form a single
strip that continues into the strips above and below.  See Appendix \ref{AppendixA}.
}
Visons are present 
at a plaquette containing two $u^{z}_{ij}$ of opposite sign, 
so we can 
set all bonds to unity, $u_{ij}=1$, 
causing the strings to vanish, establishing an equivalence
\begin{equation}\label{}
\vec{\chi}_{N} (i)\sim \vec{\chi} (i).
\end{equation}
in the axial gauge. 
The key point is that in the ground state the
majoranas $\chi_{j}$ can be treated as physical fields that describe 
the gapless spin excitations. 
We can also transform back to the original spin variables, writing the
majorana in terms of a Jordan-Wigner string
\begin{equation}\label{express}
\frac{1}{\sqrt{2}}\vec{\chi } ({j}) = \vec{S} (j) 
\left(\prod_{l \in P_{j}} \lambda^{z} (l)\right)
\times \left\{ \begin{array}{cc}
\lambda^{x} (j), &(j \in B),\cr
\lambda^{y} (j), & (j \in A),
\end{array}\right. 
\end{equation}
where the string takes a product of $\lambda^{z} (j)$
along the path $P$ consisting of sites to the left and
below site $j$ {(see appendix \ref{AppendixA})}. 

\begin{figure}[t]
\includegraphics[width=0.3 \textwidth]{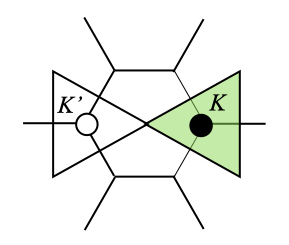}
\vspace{-0.2cm}
\caption{The hexagonal Brillouin zone of the honeycomb lattice, where
nodes of the dispersion $\epsilon (\bk )= |\gamma_{\bk }|$
lie at the vertices $ K$ and ${ K}'$.  
Majorana excitations are independently defined
over one half the Brillouin zone, described by the green 
triangle centered at ${ K}$. }
\label{Fig:BZ}
\vspace{-0.5cm}
\end{figure}

In the axial-gauge
\eqref{ksl}
becomes
\begin{equation}\label{ksl2}
H_{YL}= \frac{K}{2}\sum_{i,j } \Big[\gamma
(\bR_{i}-\bR_{j})\vec{\chi}_{A} ({i})\cdot\vec\chi_{B} ({j}) + {\rm H.c}\Big].
\end{equation}
where $\bR_{i}$ is the location of the unit cell. The hopping amplitude
$\gamma (\bR )= i  (\delta_{\bR ,0}+ \delta_{\bR,
\bR_{1}}+\delta_{\bR,\bR_{2}})$, 
where $\bR_{1,2}=(\mp \frac{\sqrt{3}}{2},\frac{3}{2})$ are the Bravais
lattice vectors.
We now Fourier
transform the Majorana fields, defining a vector majorana on each sublattice
\begin{equation}\label{twoc}
\chi_{\bk \Lambda}=\pmat{\chi^{1}_{\Lambda} (\bk ) \cr {\chi^{2}_{\Lambda}} (\bk)
\cr {\chi^{3}_{\Lambda}} (\bk)}
= \frac{1}{\sqrt{N}}\sum_{j}\pmat{
\chi^{1} _{\Lambda} (j)\cr\chi^{2} _{\Lambda} (j)\cr\chi^{3} _{\Lambda} (j)
}
e^{-i
\bk  \cdot{\bf R}_{j}}, \quad \Lambda\in (A,B),
\end{equation}
where $N$ is the number of unit cells. 
Finally, combining $\chi_{\bk A}$ and $\chi_{\bk  B}$
into a six component vector 
\begin{equation}\label{chi6}
\chi_{\bk }= \pmat{\chi_{\bk A}\cr \chi_{\bk  B}},
\end{equation}
the Hamiltonian becomes
\begin{eqnarray}\label{l}
H_{YL} = \frac{K}{2}\sum_{\bk \in \mathlarger{\hexagon}}{\chi_{\bk }}\dg 
\pmat{0 & \gamma_\bk\  \underline{1} \cr \gamma^{*}_{\bk} \ \underline{1},
& 0 }\chi_{\bk }
\end{eqnarray}
where $\gamma_{\bk }= 
 i  (1 + e^{i\bk \cdot \bR_1}+ e^{i\bk \cdot\bR_2})
$ and the momentum sum runs 
over the original hexagonal Brillouin zone ($\hexagon$). 
Now the real nature of the Majorana fermions means that 
${\chi^{a}_{\bk \Lambda}}\dg  = {\chi^{a}_{-\bk \Lambda}}$, so
the two halves of the Brillouin zone are equivalent, 
allowing us to truncate the Brillouin zone to a triangular region 
($\mathlarger{\triangleleft}$)  that surrounds the Dirac cone at
$K$ and spans
half the hexagonal Brillouin zone (see Fig.\ref{Fig:BZ}). 
In terms of this reduced Brillouin
zone, the real-space fields can be written 
\begin{equation}\label{}
\chi^{a}_{\Lambda} (j)= \frac{1}{\sqrt{N}}\sum_{\bk \in
\mathlarger{\triangleleft}}
\left(
\chi^{a}_{\bk \Lambda}e^{i\bk \cdot \bR_{j}}+ 
{\chi^{a}}\dg _{\bk \Lambda}e^{-i\bk \cdot \bR_{j}} 
 \right),
\end{equation}
and the Hamiltonian becomes 
\begin{equation}\label{honeycomb}
H_{YL}= K\sum_{\bk \in \mathlarger{\triangleleft}}{\chi_{\bk }}\dg 
( \vec{\gamma}_{\bk }\cdot\vec \alpha )\chi_{\bk },
\end{equation}
where $\vec{\gamma}_{\bk}= 
({\rm Re}\gamma_{\bk }, -{\rm
Im}\gamma_{\bk }, 0)
$ 
and the $\vec{\alpha }=
(\alpha_{1},\alpha_{2},\alpha_{3})\otimes 1_{[3]}$ are Pauli matrices
acting in sublattice space.

Diagonalizing \eqref{honeycomb} then gives 
\begin{equation}\label{YLH}
H_{YL} = \sum_{\bk \in {
\raisebox{-0.25mm}
{$\mathlarger{\triangleleft}$}
}}\epsilon_{s}({\bf k}) ( {\vec{\eta}_{\bk 1 }}\dg
\cdot{\vec{\eta}}_{\bk 1}-{\vec{\eta}_{\bk 2 }}\dg
\cdot\vec{\eta}_{\bk 2}
) , \end{equation}
where $\epsilon_{s} ({\bf k}) = K\vert \gamma_{\bk }\vert$, 
describing a Dirac cone
of majorana excitations centered at $K$.
The quasiparticle operators are given by
\begin{eqnarray}\label{l}
\vec{\eta} _{\bk 1} &=& u_{\bk }\vec{\chi }_{\bk A}+ v_{\bk
}\vec{\chi }_{\bk B},\cr
\vec{\eta } _{\bk 2}&=& -v^{*}_{\bk }\vec{\chi} _{\bk A}+
u^{*}_{\bk }\vec{\chi} _{\bk B },
\end{eqnarray}
where $u_{\bk }={1}/{\sqrt{2}}$ and $v_{\bk }= \gamma_{\bk
}/ (\sqrt{2}|\gamma_{\bk }|)$. 
In the ground-state $\vert \phi_{YL}\rangle$ all negative energy states are filled, 
\begin{equation}\label{}
\vert \phi_{YL}\rangle  =
\prod_{a\in \{1,3 \},\bk \in \mathlarger{\triangleleft} }{\eta^{a}} \dg_{\bk 2}\vert
0\rangle .
\end{equation}
The presence of a triplet of gapless ?ajoranas means that the energy cost of
visons is three times larger than in the Kitaev spin liquid, and given
by approximately $0.4K$ per vison pair\cite{Kitaev_honeycomb}.

%
%
%
%
%
%

\section{Majorana Kondo Effect}\label{sec4}

\subsection{Mean-field Hamiltonian}\label{}
We now examine the effect of the Kondo interaction on the spin
liquid.  If we rewrite this interaction in terms of the
majoranas $\vec{\chi}_{j}$, 
it becomes 
\begin{eqnarray}\label{kondoex}
H_{K}&=& J \sum_{j} (c\dg _{j}\vec{ \sigma
}c_{j})\cdot\left(-\frac{i}{2}\vec{\chi }_{j}\times \vec{\chi }_{j}\right)\cr
&= &  - \frac{J}{2} \sum_{j}c^{\dagger}_j \left[ 
\left(
\vec \sigma\cdot \vec \chi_j\right)^2 -\frac{3}{2}\right]c_j.
\end{eqnarray}
The last term in
\eqref{kondoex} can be absorbed by a shift 
in the electron chemical potential, allowing us to write
\begin{equation}\label{}
H_{K}\equiv 
 - \frac{J}{2}\sum_{j} \left(\hat  v_j^{\dagger}\hat v_j\right),
\end{equation}
where $\hat v$ is a  $S=1/2$, charge $e$ spinor boson given originally
in Eq.(\ref{vop}),
\begin{equation}\label{voppy}
\hat v_{j} =
 \bigl(\vec \sigma_{\alpha \beta }\cdot \vec \chi_j\bigr)c_{j\beta}= \begin{pmatrix}
\hat v_{j\uparrow}\\ \hat v_{j\downarrow} \end{pmatrix}.
\end{equation}
Thus the fractionalization of the spins into majoranas transforms the
Kondo interaction into an attraction that favors
the condensation of charge e spinor boson.

At temperatures low 
enough to suppress visons, there are no residual $Z_{2}$ gauge field
fluctuations, and we can consequently treat the $\hat v_{j}$ as 
a gauge neutral field. 
Taking advantage of the
bilinear form of the Kondo interaction, we now carry out a Hubbard-Stratonovich transformation
\begin{eqnarray}\label{hubbardstrat}
H_{K}&=&- \frac{J}{2} \sum_{j}c^{\dagger}_j \left[ 
\left(
\vec \sigma\cdot \vec \chi_j\right)^2\right]c_j\cr
 &\longrightarrow &\sum_{j} \biggl (\biggl[V^{\dagger}_j(\vec
 \sigma\cdot \chi_j)c_j+ {\rm
 H.c}\biggr]+2\frac{V^{\dagger}_jV_j}{J}\biggr )
\end{eqnarray}
where 
\begin{equation}\label{equiv}
V_{j}= \pmat{V_{j\uparrow}\cr V_{j\downarrow}}\equiv -\frac{J}{2}v_{j}
\end{equation}
is a spinor order
parameter. The equivalence between $V_{j}\equiv - (J/2)v_{j}$ holds at
the saddle point. 

%
%
%

It proves convenient to make a global gauge transformation on the even
(A) sublattice, $( c_{A\sigma },c_{B\sigma })\rightarrow
(-i c_{A\sigma },c_{B\sigma })$ and similarly, $(V_{A\sigma },V_{B\sigma })\rightarrow (-iV_{A\sigma },V_{B\sigma })$.
The conduction electron
Hamiltonian then takes the form 
\begin{equation}\label{HC2}
H_{C}=  -it\sum_{<i,j>} ( c\dg_{i\sigma }c_{j\sigma }-{\rm H.c}) -\mu \sum_{j}c\dg_{j\sigma }c_{j\sigma },
\end{equation}
In this gauge at $\mu=0$ the conduction and Majorana 
Hamiltonian have the same form, with opposite signs. 
Moreover, 
in the lowest energy configuration 
the Kondo bosons $\hat
v_{j}$ then conveniently condense into a uniform condensate.

The hybridization with the spin liquid induces triplet pairing,
so to proceed further, we  define
a 4-component Balian-Werthammer spinor on each sublattice 
\begin{equation}\label{}
\psi _{\bk \Lambda }=\pmat{c_{\bk \Lambda \up }\cr c_{\bk \Lambda \dw }\cr -c\dg _{-\bk \Lambda
\dw }\cr c\dg _{-\bk \Lambda\up }}, \qquad 
(\Lambda=A,B).
\end{equation}
We then merge  the two sublattice spinors
into an 8-component operator
\begin{equation}\label{psi8}
\psi _{\bk  }= \pmat{\psi _{\bk  A} \cr \psi _{\bk  B}}.
\end{equation}
In this basis, the sublattice ($\alpha $) 
charge ($\tau $) and spin ($\sigma $)  
operators are denoted by three sets of Pauli operators
given by the outer products
\begin{eqnarray}\label{l}
\vec{\alpha }_{[8]}&\equiv& \vec{\alpha }_{[2]}
\otimes 1_{[2]}\otimes 1_{[2]}
,\cr
\vec{\tau}_{[8]}&\equiv&
1_{[2]} \otimes
\vec{\tau}_{[2]} \otimes 1_{[2]},\cr  
\vec{\sigma}_{ [8]} &\equiv& 1_{[2]}\otimes
 1_{[2]} 
\otimes \vec{\sigma }_{[2]},
\end{eqnarray}
where the bracketed subscripts denoting the dimensions of the operator will be
dropped in future.  We shall use the transposed Pauli matrices for the
isospin degrees of freedom $\vec \tau_{[2]}\equiv \vec{\sigma }^{T}_{[2]}=
(\sigma_{1},-\sigma_{2},\sigma_{3})$, a choice that simplifies later
expressions. 
In this notation, 
\begin{equation}\label{hcfin}
H_{C}= \sum_{\bk \in \mathlarger{\triangleleft}}\psi\dg _{\bk } (-t\vec{
\gamma}_{\bk }\cdot\vec{\alpha}- \mu \tau_{3})\psi_{\bk }.
\end{equation}

We now introduce a four-component spinor to 
describe the Kondo hybridization, 
\begin{equation}\label{khyb}
{\cal V}_{\Lambda}=\pmat{V_{\Lambda\uparrow}\cr
V_{\Lambda\downarrow}\cr
-V^{*}_{\Lambda\downarrow }\cr
V^{*}_{\Lambda\uparrow}
}= {\rm V}_{\Lambda} {\cal Z}_{\Lambda},
\end{equation}
where 
\begin{eqnarray}\label{l}
{\cal Z}_{\Lambda}&=&\frac{1}{\sqrt{2}} \pmat{z_{\Lambda \up }\cr z_{\Lambda\dw}\cr
-z^{*}_{\Lambda\dw }\cr z^{*}_{\Lambda\up}}\\ \phantom{\int} \nonumber
\end{eqnarray}
is normalized to unity ${\cal Z}_{\Lambda}\dg  {\cal Z}_{\Lambda}=1$, and we use a
Roman V$_{\Lambda}$ to denote the magnitude of the hybridization. 
The hybridized Kondo term then becomes
\begin{equation}\label{KLhyb}
H_{K }= \sum_{ \Lambda=A,B}
\left( 
\sum_{\bk \in \mathlarger{\triangleleft}} \biggl[
(\psi \dg_{\bk \Lambda}\vec{\sigma }{\cal V}_{\Lambda})\cdot
\vec{\chi}_{\bk\Lambda}
+
{\rm H.c}
\biggr]+2N\frac{{\rm V} _{\Lambda}^{2}}{J}
\right).
\end{equation}
We shall focus on the uniform case, where ${\cal V}_{A}={\cal V}_{B}={\cal V}$, which
creates the largest hybridization gap. 
Combining \eqref{hcfin} \eqref{KLhyb} and \eqref{honeycomb}, the 
mean-field Hamiltonian can be compactly written as 
\begin{equation}\label{Hfull}
H = \sum_{\bk  \in \mathlarger{\triangleleft}}\Psi\dg_{\bk }\pmat{ - t
(\vec{\gamma}_{\bk }\cdot \vec{\alpha}) - \mu\tau_{3}&
\vec\sigma
{\cal V}
\cr
{\cal V}\dg 
\vec\sigma
& 
K (\vec{\gamma}_{\bk }\cdot \vec{\alpha}) }\Psi_{\bk } + 4 N
\frac{{\rm V}^{2}}{J}
\end{equation}
where 
\begin{equation}\label{}
\Psi_{\bk }= \pmat{\psi_{\bk }\cr
\chi_{\bk }}, 
\end{equation}
where  $\psi_{\bk }$ is defined in \eqref{psi8} and $\chi_{\bk }$ is
defined in \eqref{chi6}. In the off-diagonal components of \eqref{Hfull}
we have used the short-hand 
$\psi \dg_{\bk }\vec{\sigma }{\cal V} {\chi}_{\bk }
\equiv \psi \dg_{\bk } (\vec{\sigma }\cdot \vec{\chi}_{\bk }){\cal V}
$
and 
$\vec{\chi}\dg _{\bk }{\cal V}\dg \vec{\sigma }\psi_{\bk }\equiv
{\cal V}\dg (
{\vec{\chi }}\dg _{\bk }\cdot\vec{\sigma })\psi_{\bk }
$. 
\fg{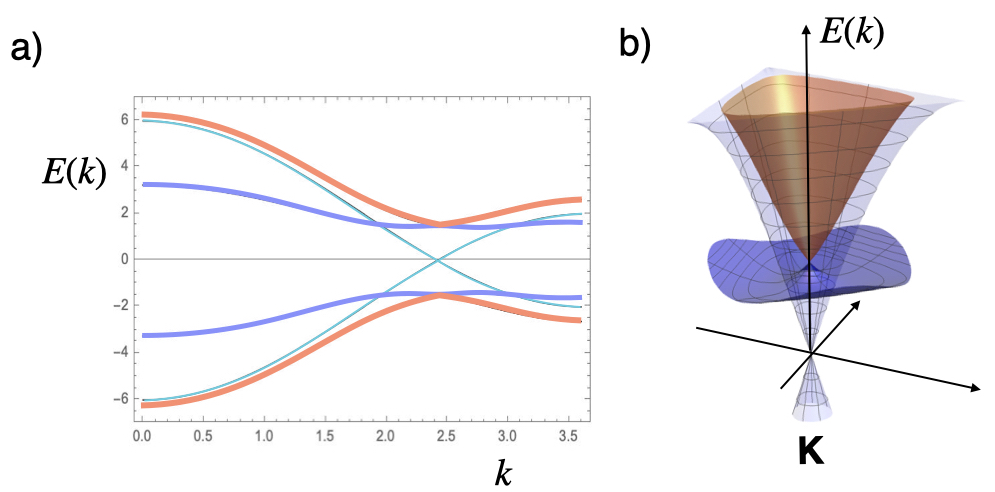}{spec1}{a) Spectrum of the Kitaev Kondo lattice for $\mu=0$, 
taking units $K=1$, $t=2K$ and ${\rm
V}=1.5K$. Thick orange lines denote gapped conduction electron lines,
thick blue lines denote the gapped Majorana spin excitations
while while thin blue lines denote the single gapless Majorana 
conduction band. b) Schematic three dimensional plot of spectrum in the vicinity
of the $K$ point. }

\subsection{The case of half filling ($\mu=0$).}\label{}

If we  now split the conduction sea into scalar and vector components
\begin{eqnarray}\label{split}
{\psi}^{0}_{\bk \Lambda}&=&{\cal Z}\dg\cdot\psi_{\bk \Lambda},\cr
\vec{\psi}_{\bk \Lambda}
 &=& {\cal Z}\dg\cdot\vec{\sigma }\cdot\psi_{\bk \Lambda },
\end{eqnarray}
then from \eqref{KLhyb}
we can decouple
\begin{equation}\label{decup}
(\psi \dg_{\bk \Lambda}\vec{\sigma }{\cal V})\cdot
\vec{\chi}_{\bk\Lambda}= 
{\rm V}(\psi \dg_{\bk \Lambda}\vec{\sigma }{\cal Z})\cdot
\vec{\chi}_{\bk\Lambda}= {\rm V} (\vec{\psi}\dg_{\bk \Lambda} \cdot \vec{\chi
}_{\bk \Lambda}).
\end{equation}
In other words, only the vector Majorana components of the conduction
sea hybridize with the spin liquid, and the scalar part is unhybridized.
At particle-hole symmetry $\mu=0$, the Hamiltonian decouples
into a gapless scalar conduction sea and gapped
vector sea of excitations
\begin{eqnarray}\label{l}
H&=& \sum_{\bk   \in \mathlarger{\triangleleft}}\psi \dg_{0\bk } (-t
 \vec{\gamma}_{\bk }
\cdot\vec{\alpha }
)\psi_{0\bk }+ 4N\frac{{\rm V}^{2}}{J}
 \cr
&+& \sum_{\bk   \in
\mathlarger{\triangleleft}}
\pmat{\vec{\psi}\dg_{\bk }& \vec{\chi }\dg _{\bk }}
\pmat{- t \vec{\gamma}_{\bk }\cdot \vec{\alpha} &
{\rm V}\cr {\rm V}&  K \vec{\gamma}_{\bk }\cdot \vec{\alpha} 
}\pmat{\vec{\psi }_{\bk }\cr \vec{\chi}_{\bk}}.
\end{eqnarray}
The fourteen eigenvalues of this Hamiltonian involve a single Dirac cone 
with the two eigenvalues $\pm \epsilon_{c} (\bk )$ of 
the original conduction sea (light blue curve in Fig. \ref{spec1}), 
and  four triply degenerate gapped excitations (blue (+) and red (-) curves in
Fig. \ref{spec1})
with eigenvalues $\pm E^{+}_{\bk }$ and $\pm E^{-}_{\bk }$, where
\begin{equation}\label{}
E^{\pm }_{\bk } = 
\sqrt{{\rm V}^{2}+ \biggl(\frac{\epsilon_{c} (\bk )+\epsilon_{s} (\bk )}{2}\biggr )^{2}
}\pm \biggl ( \frac{\epsilon_{c} (\bk )-\epsilon_{s} (\bk )}{2}\biggr ).
\end{equation}
Here $\epsilon_c ({\bk })= t\vert
\gamma_{\bk }\vert $ and $\epsilon_{s} (\bk )= K |\gamma_{\bk
}|$\eqref{YLH}.

Figure \ref{spec1}. shows a representative spectrum.  The Dirac conduction band
is composed of four degenerate majoranas: three of these hybridize with the spin
liquid, pushing the Dirac cone intersection to a finite energy $V$,
while the fourth Majorana component decouples from the spin liquid
as a single gapless Dirac cone.

From these dispersions, we can calculate the mean field Free energy to be
\begin{eqnarray}\label{l}
F[T] &=& - T \sum_{\bk   \in
\mathlarger{\hexagon}}
 \ln \left[ 2 \cosh \left(\frac{\beta \epsilon_{\bk
}}{2} \right)\right]\cr
&-&  3T \sum_{\bk   \in
\mathlarger{\hexagon}, \pm }
 \ln \left[ 2 \cosh \left(\frac{\beta E^{\pm}_{\bk
}}{2} \right)\right] + 4 N \frac{{\rm V}^{2}}{J}
\end{eqnarray}
so the ground-state energy per unit cell is 
\begin{equation}\label{}
\frac{E}{N} = 
- \frac{A_{c}}{2}
\int_{\bk   \in
\mathlarger{\hexagon}} \frac{d^{2}k}{(2\pi)^{2}}
\biggl(\epsilon_{\bk} +3 (
 E^{+}_{\bk}+ 
 E^{-}_{\bk})\biggr)
 + 4  \frac{{\rm V}^{2}}{J}
\end{equation}
where $A_{c}=\frac{3\sqrt{3}}{2}$ is the area of the unit cell.

Differentiating with respect to ${\rm V}^{2}$ leads to a gap equation
\begin{equation}\label{}
 \frac{A_{c}}{2} \int_{\bk   \in
\mathlarger{\hexagon}} \frac{d^{2}k}{(2\pi)^{2}}
\frac{1}{\sqrt{{\rm V}^{2}+
(\epsilon_{s} (\bk )+\epsilon_{c} (\bk ))^{2}/4
}}
= \frac{4}{3J}
\end{equation}
If we introduce the scaled quantities $g = J/ [3(t+K)]$ (note: $3t$
and $3K$ are the half-band widths of the conduction and Majorana
bands, respectively) and $v={\rm V}/ (t+K)$,
then the mean-field equation for the gap becomes 
\begin{equation}\label{mft}
\frac{1}{g} = \Phi (v)
\end{equation}
where 
\begin{equation}\label{mft2}
\Phi (v) = \frac{9A_{c}}{4} \int_{\bk   \in
\mathlarger{\hexagon}} \frac{d^{2}k}{(2\pi)^{2}}\frac{1}{\sqrt{4v^{2}+
|\gamma_{\bk }|^{2}}}.
\end{equation}
A quantum critical point separating spin-liquid/metal from the 
an order fractionalized
phase is located at $g_{c}= 1/\Phi (0)= 0.5$. 
Fig. \ref{mfp} shows a plot of the hybridization versus coupling constant
predicted by the mean-field theory.
\fg{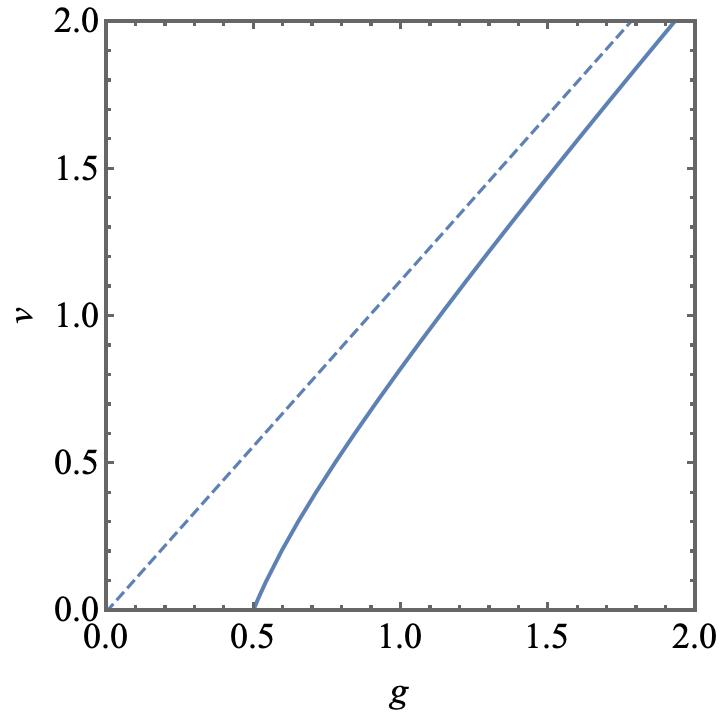}{mfp}{Plot of $v=V/ (t+K)$ 
vs $g=J/ 3(t+K)$ predicted by equations
\eqref{mft} and \eqref{mft2}. { The dashed line gives the asymptotic
large $g$ limit $\nu=(9/8)g$ of the phase boundary.} 
}

\section{Vicinity of the QCP }\label{sec5}

The vicinity to the quantum critical point at $g=g_{c}$, $\mu=0$ is of
particular interest.  Phase transitions in 
systems with Dirac spectrum lie in the class of of Gross-Neveu-Yukawa
models \cite{boyack2004}. 
Renormalization group analyses of this class of models indicate that 
the quantum critical point acquires full  Lorentz invariance (which in
mean-field theory corresponds to the case $t=K$). 
The corresponding long-wavelength action 
for  our case is the deformation of Eq.(\ref{Hfull}):
\begin{eqnarray}\label{Hfull2}
&& H = \sum_{\bk  \in \mathlarger{\triangleleft}}\Psi\pmat{ - it
(\vec{\alpha}\cdot \vec\nabla
) - \mu\tau_{3}&
\vec\sigma
{\cal V}
\cr
{\cal V}\dg 
\vec\sigma
& 
 it(\vec{\alpha }\cdot \vec\nabla) }\Psi + \nonumber\\
 && \frac{1}{2}{\cal V}^+(-\nabla^2 + m^2){\cal V} + \lambda({\cal V}{\cal V}^+)^2. 
\end{eqnarray}
The survival of a relativistic majorana in the broken symmetry phase
is rather striking consequence of the 
mismatch between the number of electron and majorana channels. These
features may be of interest in the generalization of these ideas from
quantum materials to exotic scenarios of broken symmetry in the
vacuum. 
\fgb{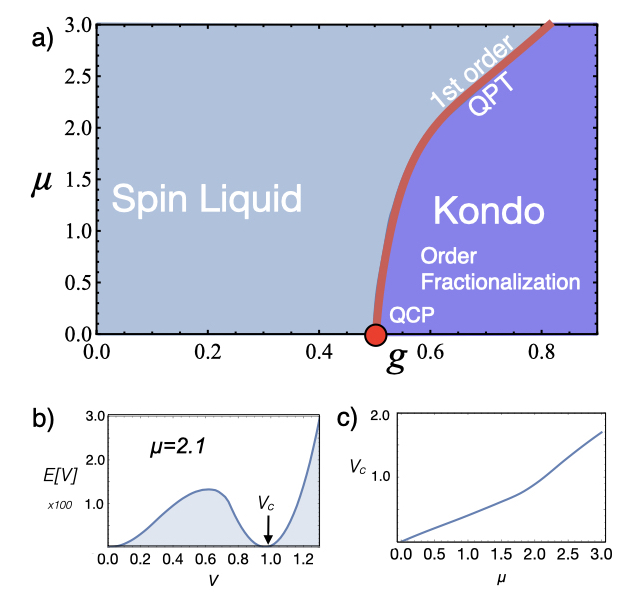}{pdia}{a) {Mean field} phase diagram as a function of chemical
potential $\mu$. At $\mu=0$, the quantum phase transition from the spin liquid to
the Kondo phase is a quantum critical point. b) representative plot of mean-field energy $E[V]$ versus $V$
illustrating first order minimum that develops at finite 
$\mu$. c) Critical value $V_{c}$ at first
order quantum phase transition, showing that $V_{c}/\mu\sim 0.5$,
corresponding to a direct transition into a state with a single
neutral majorana cone of excitations. Calculations were made using $K=1,
t=2$. }

\subsection{Finite doping  $\mu\neq  0$}\label{}

At finite doping, away from charge neutrality, 
the decoupled conduction sea develops a Fermi
surface. In principle, the excitation spectrum of the condensate
becomes more complex, for at
$V=0^{+}$,  there are three Fermi surfaces: two derived from the
conduction electrons and one derived from the $\chi $ fermions.
In principle, as the hybridization is increased from zero, 
the Fermi surfaces undergo a sequence
of Lifschitz transitions, until entirely disappearing once $|V|\sim |\mu|/2$,
they entirely vanish, entering the phase with a single
neutral Majorana excitation cone, qualitatively identical to that obtained 
at $\mu=0$. (See Fig. \ref{pdia}.).

However, the mean-field theory predicts that these intermediate
phases are entirely bypassed by a first order
transition into the high $V$ state. 
This can be {qualitatively} understood from the dependence of the 
Ginzburg Landau free energy on $\mu$ and $V$, which in the
vicinity of the $\mu=0$ QCP 
is given by 
\begin{equation}
E = a\mu^2 V + \tau V^{2} + b V^{3},
\end{equation}
where {$\tau \propto  (g_{c}-g)$}. The coefficients $a$ and
$b$ depend on the ratio
 $t/K$,{\  and can be evaluated explicitly for the relativistic 
case $t/K=1$, confirming that they are both positive}. 
The  term linear in $V$ results 
from the development of a gap  $V$  in the 
electron Dirac cones.  Since the density of states
 is proportional to energy, in the normal state there is a Fermi
 surface containg $O (\mu^{2})$ electrons, giving rise to an increase in energy of order
$\mu^{2}|V|${, so that $a>0$.} This energy is reminiscent of the
Van der Waals equation of  state in the vicinity of the liquid-gas
critical point, and gives rise to a first order phase transition
at $\tau = -2 |\mu| \sqrt{ab}$ into a state with $V_{c}=|\mu|\sqrt{\frac{a}{b}}$.
Fig. \ref{pdia} displays a detailed calculation
of the mean-field free energy at finite doping, showing that at the
crical $g$, a minimum in the free energy degenerate with the ground-state
develops at finite $V_{c}\sim 0.5|\mu|$.

\section{Fractionalized order}\label{sec6}

We now address the nature of the long-range order associated with the
hybridization between electrons and Majorana fermions. Since the
hybridization carries $Z_{2}$ gauge charge, the
definition of long-range order requires the insertion of a gauge 
string.  We can construct the following density matrix 
\begin{eqnarray}\label{l}
{\Sigma }_{ab} (x,y )&=&
\langle  \hat {\cal V}_{a} (x)
\hat \S (x,y) \hat { \cal V}\dg _{b} ({y})\rangle, 
\end{eqnarray}
where $\hat \S (x,y) = \prod_{l}u_{(l+1,l)}$
is the string operator linking the sites $x $ and $y $
and ${\cal V} (x )$ is the four-component Kondo hybridization
introduced in  \eqref{khyb}
\begin{equation}\label{}
\hat {\cal V} (x ) = \frac{1}{\sqrt{2}}\pmat{v_{\uparrow}(x )\cr v_{\downarrow } (x )
\cr - v\dg_{\downarrow } (x )\cr v\dg_{\uparrow } (x )}.
\end{equation}
$\Sigma_{ab}(x,y)$ determines the 
amplitude for an electron to coherently tunnel through the spin
liquid from $y$ to $x $. 

Like the underlying gauge fields, 
the gauge strings $\hat {\cal P} (x,y )$
are {\sl constants of motion}, commuting with the
Hamiltonian and the constraints. 
The energetic cost of visons allows us to 
safely set 
all $u_{i_{A},j_{B}}=1$ in the ground-state, so
the string variable is simply unity,
\begin{equation}\label{}
\hat  \S (x,y )=1, \qquad \hbox{(axial gauge)},
\end{equation}
and in this gauge, $\Sigma (x ,y )$ reverts to a 
conventional two point functions.
This is precisely the gauge we have used for the mean-field theory, 
so the mean-field density matrices 
factorize
into a product of the spinors
\begin{eqnarray}\label{l}
\Sigma_{ab} (x,y)= {\cal V}_{a} (x) {\cal V}\dg _{b}(y).
\end{eqnarray}
Importantly, since this factorization occurs in
a $Z_{2}$ {\sl gauge invariant} quantity, it is true in all gauges,
and is thus
immune to the 
average over gauge orbits that annihilates gauge dependent quantities
(the origin of Elitzur's no-go theorem). 
Of course,  mean-field theory is 
corrected by Gaussian fluctuations of the fields, but 
these $Z_{2}$ gauge invariant corrections are no different to the corrections
that occur in conventional ODRLO, so we expect
that beyond a coherence
length, the factorization  will be preserved 
as an asymptotic long-distance property, 
i.e 
\begin{eqnarray}\label{l}
\langle  {\cal V} ({x})\hat \S (x,y){\cal V}\dg ({y})\rangle 
\xrightarrow[]{|x-y|\rightarrow
\infty}
{\cal V} (x ){\cal V}\dg  (y ).
\end{eqnarray}
This is the phenomenon of order fractionalization. 

We can extract two interesting quantities from this density matrix, a
$Z_{2}$ string-expectation value
\begin{equation}\label{z2string}
Z (x,y)
= {\rm Tr}[\Sigma (y,x) ] = \langle \hat {\cal  V}\dg
(x)
\hat {\cal P} (x,y)\hat {\cal V} (y)
  \rangle, 
\end{equation}
and an $SO (3)$ matrix
\begin{eqnarray}\label{}
D_{ab} (x,y)& =& {\rm Tr}[\Sigma (y,x) \sigma^{a}\tau^{b}]\cr
 &=& \langle \hat {\cal V}\dg
(x)\sigma^{a}\tau^{b}\hat {\cal V} (y)  
\hat {\cal P} (x,y)
\rangle, 
\end{eqnarray}
The local density matrix $D_{ab}(x ,x ) = (2V^{2}/J) [\hat  d^{b}
(x)]_{a}$ determines the composite magnetism
and pairing at site $x$, where the $(\hat d^{b})_{a}= {\cal Z}\dg
(x)\sigma^{a}\tau^{b}{\cal  Z} (x)$ is the triad of local vectors
introduced in \eqref{triad} and we have normalized the spinors using
\eqref{equiv} and \eqref{khyb}. 
The composite order $D_{ab} (x,x)$ only determines the spinor-order up
to a sign. 
However, the factorization of the scalar $Z (x,y)$ 
\begin{equation}\label{}
Z (x,y) \xrightarrow[]{|x-y|\rightarrow
\infty}{\cal V}\dg (x){\cal V} (y)
\end{equation}
is sensitive to the relative sign of the order parameter $v (x )$ 
at sites $x $ and $y $. 

We can further emphasize the physical nature of these
results by rewriting the order parameter and
the gauge string in terms of  variables from the original
model. Using \eqref{vop} and the constraint $-2i\Phi_{j}^{S}\Phi_{j}^{T}=1$\eqref{constraint}
we can rewrite the hybridization field in terms of the composite operator 
${\cal F}_{j}=
(\vec{ \sigma
}\cdot
\vec S_j)c_{j}
$,
as 
\begin{eqnarray}\label{replace}
\hat  v_{j }= (\vec{ \sigma }\cdot\vec\chi_j)c_{j}= 2i \Phi^{T}_{j}{\cal F}_{j}.
\end{eqnarray}
By substituting $u_{(i,j)}=-2i
\epsilon_{x_{i}x_{j}}b^{\alpha_{ij}}_{i}b^{\alpha_{ij}}_{j}$, where
\begin{equation}\label{epsy}
\epsilon_{x',x}\equiv \pmat{1& 1\cr -1 & -1}
=
\left\{ 
 \begin{array}{cc}
1 & (x'\in A)\cr
-1 & (x' \in B)
\end{array}\right.
\end{equation}
into the string $\hat \S (x',x) = \prod_{l}u_{(l+1,l)}$, then using the relation
$\vec{\lambda_{j}}= -i \vec{b}_{j}\times
\vec{b}_{j}$ \eqref{key} at the bond intersections
and 
$b^{\alpha }_{j}=\Phi_{j}^{T}\lambda_{j}^{\alpha }$ \eqref{defined} 
at its two ends, 
we can 
rewrite the string as  $\hat  \S=- 2i \Phi^{T}_{x'}\Phi^{T }
_{x}\S'$, where 
\begin{equation}\label{unique}
\hat \S '(x',x) = s (x',x)
\prod_{j\in {\cal P}} \lambda_{j}^{a_{j}}
\end{equation}
is a product of $\lambda$ operators taken along
directions $a_{j}$ extremal to the path $\cal P$ (see Fig. \ref{figk}), including
an initial $\lambda^{a_{i}}_{i}$ and final $\lambda^{a_{f}}_{f}$
operator, oriented along the initial and final bonds.
The parity $s(x',x)=\pm 1$ 
is determined
by the relative
directions of the initial 
and final bond-vectors, $\hat v$ and $\hat v'$
\begin{equation}\label{}
s (x',x) =
\left\{ 
 \begin{array}{cl}
\epsilon_{x',x},& (\hat v\cdot \hat v'=0)
\cr
 {\rm sgn}[
 (\hat  v'\times \hat  v)\cdot\hat  z],& (\hat v\cdot \hat v'\ne 0)
\end{array}\right.
\end{equation}
where $\hat z$ is normal to the plane.

In the ground-state,
${\cal P}$ does not depend on the path, so 

\figwidth=0.8\columnwidth
\fgh{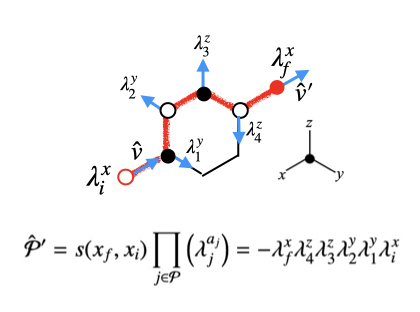}{figk}{An example of the string operator ${\cal
P}'$
in Eq.(\ref{unique}). }

\begin{eqnarray}\label{l}
-2i\langle {\cal F}_{a} (x )
{\cal F}\dg _{b} (y )
\hat {\cal P}' (x ,y )
 \rangle 
\xrightarrow[]{|x-y|\rightarrow
\infty}
v_{a} (x) v\dg_{b} (y). 
\end{eqnarray}
In other words, the composite fermions have developed a gauged off-diagonal long-range
order. 
By \eqref{replace}, the composite fermions 
split up into a bosonic spinor with long range
order, and an ancillary majorana that 
decouples from the Hilbert space:
\begin{equation}\label{}
{\cal F}_{j}= (\vec{ \sigma }\cdot\vec S_{j})c_{j}= -i v (x )
\Phi^{T} (x).
\end{equation}
This remarkable transmutation in the statistics of the composite
fermions is a direct consequence of order fractionalization.

\figwidth=0.7\columnwidth
\fgh{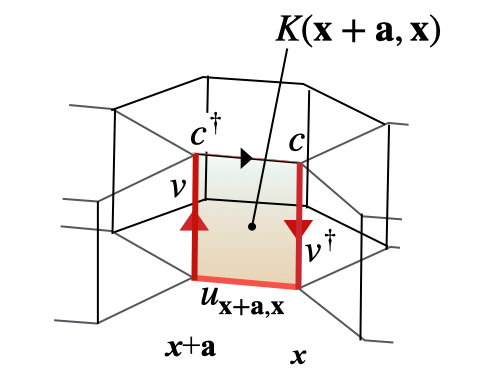}{figk}{Kondo plaquet. The energetic cost of flux
through the blue area favors a uniform arrangement of the spinor order
parameter. }

\subsection{Topology and Vison Confinement}\label{}

The Yao Lee and Kitaev spin liquids 
have a topological degeneracy. We now discuss how this is modified by
the presence of a Kondo hybridization. 
Suppose we have a domain across which 
the $Z_{2}$ string \eqref{z2string}
$Z (x,y)$ changes sign. 
At the domain boundary,
there is a ``Kondo flux'' identified with an inter-layer  plaquette that links
the conduction and spin fluids
\begin{equation}\label{}
K (x +{\bf a},x ) = \langle  
c \dg (x +{\bf a})
v (x +{\bf a})u_{x +{\bf a},x }v\dg  (x )c(x )\rangle.
\end{equation}
$K (x+{\bf a},x)$  describes the amplitude for an electron  to traverse a rectangular
path entering the spin liquid at $x $, exiting at $x+{\bf a}$ and
returning via the conduction sea (see Fig \ref{figk}). 
This favors a ground-state with a spinor $v (x )$ that
is uniform in both direction and sign.

If we separate two visons without allowing the spinor background to deform, then 
we create a ladder of bonds with $u_{ij}=-1$.   The reversed sign in
the ladder then gives rise to a Kondo domain wall with 
cost $E (L)\propto L$ proportional to its length.  In this situation, the
visions would be linearly confined, and the cost of a $Z_{2}$ vortex
through the torus would be proportional to the circumference of the
torus, $E_{Z_{2}}\propto L_{C}$ (Fig. \ref{figdom}).  This is the situation we would
expect if the Kondo hybridization were a $Z_{2}$ scalar (a situation
that would occur if the conduction band were made of three, rather
than four majoranas, giving rise to an O(3), rather than an SU(2)
Kondo model.)

However, 
the two visons can remove the domain wall by binding
themselves to a ``$2\pi$'' or ``$h/2e$'' vortex in which the principle axes $\hat d^{b}$ rotate
about some axis through $2\pi$.  A $2\pi$ rotation of the spinor 
${\cal V} (\theta )= e^{-i \frac{\theta }{2}\tau_{z}}{\cal V}_{0}$ 
causes it to pick up a minus sign: ${\cal V} (2\pi)= - {\cal V}_{0}$. 
In isolation, such a vortex would give rise to a  sharp
discontinuity in the spinor, but if the jump in ${\cal
V}$ is located along the ladder where the gauge field $u_{ij}$ changes
sign, then the Kondo domain is now removed and the
gauged density matrix $\Sigma_{ab} (x,y)$ remains a smooth, single
valued function.   Since the energy cost of separating two vortices grows
as $\ln L$, the resulting vortex-vison bound-state is logarithmically
confined. 

\figwidth=\columnwidth
\fg{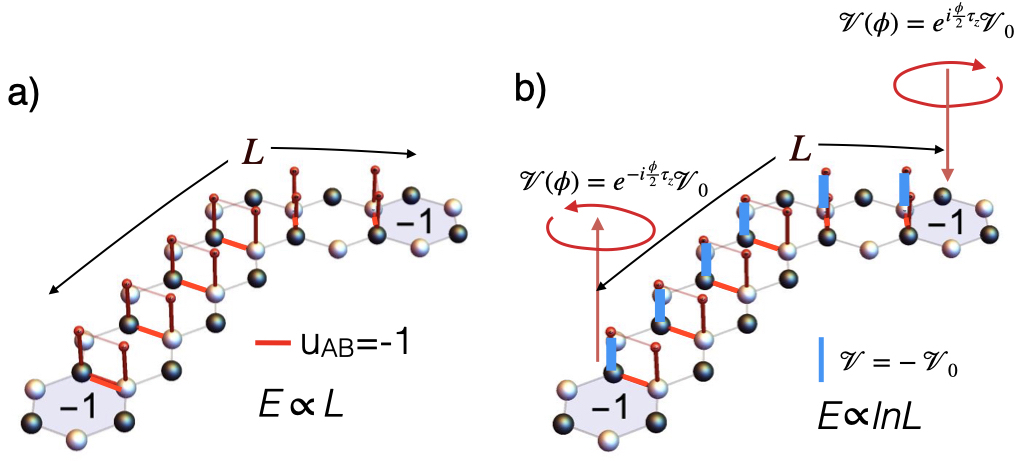}{figdom}{a) When two visons are separated a
distance $L$, they
introduce a ladder of $u_{ij}=-1$, giving rise to Kondo domain wall
whose energy grows linearly with length $E (L)\propto L$.
(b) If each vison bind to a ``$2\pi$'' vortex of the ${\cal V}$ field,
the domain wall is eliminated, so the vison-vortex combination is
logarithmically confined, with an energy $E (L)\propto \ln L$.
}

\figwidth=\columnwidth
\fgb{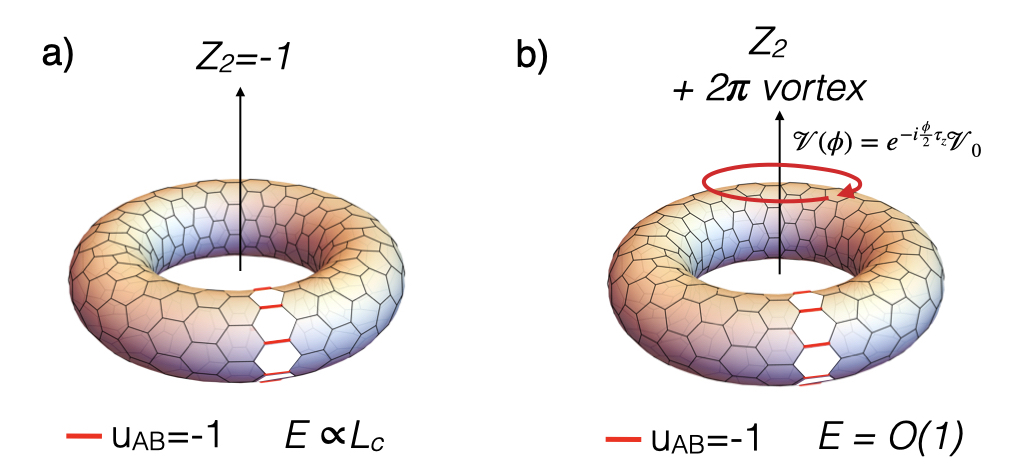}{figdom}{a) A $Z_{2}$ vortex through the
torus is formed by a row of $u=-1$ bonds.  In the Yao-Lee model, the
domain wall costs no energy, but in the presence of a uniform order
parameter $\cal V_{0}$, the domain wall costs an energy $E (L)\propto
L_{C}$, where $L_{C}$ is the circumference of the torus. 
(b) When the $Z_{2}$ vortex binds a $2 \pi$ vortex of the hybridization,
the domain wall is removed, giving rise to an energy $E\sim O (1)$
which is intensive in the torus dimensions, restoring 
the single-valued character of the gauged off-diagonal
long-range order.
}

In a similar fashion, if we create two visons, separating them and 
re-annihilate them after passing one around a ring of
the torus, we create a $Z_{2}$ vortex through the torus, with a 
Kondo domain wall that passes right around it ( see Fig. \ref{figdom}a).  This process costs no
energy in a pure Kitaev or Yao Lee model, and is the origin of their
topological degeneracy. Let us now consider the influence of the Kondo
effect. In a model where the
Kondo coupling were $Z_{2}$ scalar\cite{colemanioffetsvelik} { as in the double layer Yao-Lee model with the Heisenberg exchange interaction between the layers},  this would cost an energy
proportional to the circumference $L_{C}$  of the torus.
However,  if we also
introduce a vortex in which the ${\cal V} (x)$ rotates through $2\pi$,
as shown on Fig. \ref{figdom}b, the spinor picks up
an additional minus sign in passing around the vortex, and we then
remove the discontinuity in the function $\Sigma _{ab} (x,y)$,
removing the domain wall.  This state with a combined $Z_{2}$ and $2\pi$ vortex only
costs the energy to twist the spinor order through $\pi$, which
involves an elastic energy $(\rho_s/2)  (\pi/L)^{2} \times LL_C=\pi^2 \rho_s /2(L_C/L)$, where
$\rho$ is the stiffness, a value which is 
intensive in the linear size.  
Since this configuration cannot be smoothly
returned to the original ground-state without creating a Kondo domain
wall, the bound combination of a $2 \pi$ vortice and vison pairs { is
topologically distinct excitation of the  ground-state.}

\section{Triplet pairing condensate}\label{sec7}

\subsection{Electron self-energy}\label{}
\fgb{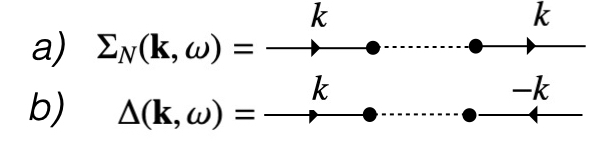}{feyn}{When an electron scatters through the spin
liquid it can emerge a) as an electron, giving rise to resonant
scattering  and b) as a hole, giving rise to resonant Andreev
reflection. }
Once the spinor $\hat v_{j}$ condenses (dark blue region of Fig. \ref{pdia}),
the resulting condensate 
will coherently scatter electrons through the spin liquid.  
The order fractionalization means that an electron can remain
submerged within the 
spin liquid over arbitrarily long distances. 
When
a Majorana spinon resurfaces into the electron fluid, 
it can do so as either a particle, or a hole, 
so the scattering amplitude of electrons
via the spin liquid develops both normal and anomalous (Andreev) scattering 
components (Fig. \ref{feyn}). To examine these processes, consider 
the conduction electron self-energy that
results from integrating out the majoranas $\chi_{\bk }$.
{
From \eqref{honeycomb}, the propagator for the majoranas
in the unhybridized spin liquid is 
\begin{equation}\label{}
G^{(0)}_{\chi } (\bk ,\omega) = \frac{1}{\omega -  {\vec{\gamma}_{s\bk }}\cdot \vec{\alpha }}\ ,
\end{equation}
where  $\vec{{\gamma}_{\bk }}= ({\rm Re}\gamma_{\bk }, -{\rm
Im} \gamma_{\bk }, 0)$ and 
$
\gamma_{s\bk }= K \gamma_{\bk } =  i  K(1 + e^{i\bk \cdot \bR_1}+ e^{i\bk \cdot\bR_2})
$. 
Integrating out the majoranas in \eqref{Hfull} then  introduces a
self-energy to the conduction electrons given by 
\begin{eqnarray}\label{}
\Sigma (\bk ,\omega)&=&
\sigma^{a}{\cal  V}G^{(0)}_{\chi } (\bk ,\omega){{\cal V}\dg }\sigma^{a}\cr
&=&
\sigma^{a}{\cal  V}
\frac{1}{\omega -  \vec{\gamma}_{s\bk }\cdot \vec{\alpha }
}{{\cal V}\dg }\sigma^{a},
\end{eqnarray}
} 
where 
we sum over the  repeated index $a$.

By commuting the  ${\cal V}= {\rm V}{\cal Z}$
through the $\vec{\alpha} $, we obtain
\begin{eqnarray}\label{l}
\Sigma (\bk ,\omega) =
\frac{{\rm V}^{2}}{\omega - \vec{\gamma}_{s\bk }\cdot \vec{\alpha }} (\sigma^{a}{\cal Z}{\cal Z\dg }\sigma^{a}).
\end{eqnarray}
Using the identity ${\cal Z}{\cal Z}\dg + 
\sigma^{a }{\cal Z}{\cal Z}\dg \sigma^{a} = 1 
$, we can then write the conduction self-energy in the form 
\begin{equation}\label{}
\Sigma (\bk ,\omega) = 
\frac{\rm V^{2}}{\omega - \vec{\gamma}_{s\bk }\cdot \vec{\alpha }} (1-{ P}), 
\end{equation}
where ${P}= {\cal Z}{\cal Z}\dg$ projects onto the zeroth
Majorana component of the conduction sea, which consequently 
does not hybridize with the spin liquid. 
Without the projector ${ P}$, this scattering would describe a Kondo
insulator on a honeycomb lattice: the introduction of the 
projector breaks both time-reversal
and gauge symmetry by eliminating a specific Majorana component of the
conduction sea. 

To examine the pairing components of the self-energy we write
${\cal Z}{\cal Z}\dg = 
\frac{1}{4} 
(1+ d_{ab}\sigma^{a}\tau^{b})
$
where the $d_{ab}\equiv ({\bf d}^{b})_{a}= {\cal Z}\dg  \sigma^{a
}\tau^{b}{\cal Z}$
are the triad of orthogonal vectors $ ({\bf d}^{1}, {\bf
d}^{2}, {\bf d}^{3})$ that define the composite SO(3) order (see \eqref{triad}).  For the choice  ${\cal Z} = {\cal
Z}_{0}=
\frac{1}{\sqrt{2}} (1,0,0,1)^{T}$,
the d-vectors
align with the co-ordinate axes, $({\bf d}^{1}, {\bf d}^{2}, {\bf d}^{3}) = (\bx ,\by ,{\bf z} )$.  The 
resonant scattering 
off the  spin liquid takes the form 
\begin{equation}\label{}
\Sigma (\bk ,\omega) 
= {\rm V}^{2} 
\tfrac{1}{4}
\left(3 - ({\bf d}^{b}\cdot\boldgreek{\sigma} )\tau^{b} \right)
\frac{\omega +  \vec\gamma_{s\bk }\cdot \vec{\alpha } }
{\omega^{2} - |\gamma_{s\bk }|^{2}}.
\end{equation}
We can divide the self-energy 
into normal and pairing components 
\begin{equation}\label{}
\Sigma = \Sigma_{N}+\Delta (\bk ,\omega)\tau_{+}+ \Delta \dg (\bk
,\omega)\tau_{-},
\end{equation}
where
\begin{eqnarray}\label{comps}
\Sigma_{N} (\bk ,\omega)&=&
\tfrac{1}{4}
\left(3 - ({\bf d}^{3}\cdot \boldgreek{ \sigma })\tau_{3}  \right)\Sigma_{0} (\bk
,\omega),\cr
\Delta ( {\bk ,\omega})&=& -
\tfrac{1}{4}\left (
({\bf d}^{1}+i {\bf d}^{2})\cdot \boldgreek{\sigma } \right)\Sigma_{0} (\bk
,\omega).
\end{eqnarray}
Here,
\begin{equation}\label{}
\Sigma_{0} = {{\rm V} }^{2}
\left(
\frac{\omega +  \vec\gamma_{s\bk }\cdot \vec{\alpha } }
{\omega^{2} - |\gamma_{s\bk }|^{2}}
\right).
\end{equation}
$\Sigma_{N}$ describes a kind of odd-frequency
magnetism (with no onsite magnetic polarization).  The second-term
$\Delta (\bk ,\omega)$ in \eqref{comps}
describes a triplet gap function,
with a complex d-vector 
$\hat 
{\bf d}^{1}+ i {\bf d}^{2}$ which breaks time-reversal
symmetry. 

The frequency, momentum and
sublattice structure of 
the pairing is an interesting illustration of the
 SPOT=-1\cite{sasha19} acronym
for the exchange-antisymmetry of
pairing,  where S, P, O and T are the
parities of the pairing under spin exchange, spatial inversion,  sublattice 
exchange and time inversion, respectively.  Here, since  the pairing is
triplet and spin symmetric (S=1), POT=-1: there are in fact 
three separate odd-frequency, odd-parity and odd-sublattice components. 
The term proportional to $\omega$ is odd frequency, 
(T=-1, P=O=+1), while the even-frequency component (T=+1)
divides into two parts
$
\vec{\gamma}_{s\bk }\cdot \vec{\alpha }= 
(\gamma^{1}_{s\bk }\alpha_{1}+ \gamma^{2}_{s\bk }\alpha_{2}) 
$ which are respectively,  odd parity, sublattice even (P=-1, O=+1)
and even parity, sublattice odd (P=+1, O=-1). 

\subsection{Long-range tunneling}\label{}

The structure of the self-energy reflects the long-range tunneling of
electrons through the spin liquid.  If the order parameter ${\cal
V} (x)$ varies slowly in space and time, the electron self-energy
takes the form 
\begin{equation}\label{}
\Sigma (x,x')= \sigma^{a}{\cal V} (x){\cal G} (x-x'){\cal V}\dg (x')\sigma^{a}
\end{equation}
where
\begin{equation}\label{}
{\cal G} (x-x') = \int \frac{{}d^{3}k}{(2\pi)^{3}} 
\left(\frac{1}{\omega -
\gamma_{s\bk }\cdot \vec{\alpha }} 
 \right)e^{i (\bk \cdot \bx-\omega t)}
\end{equation}
is the majorana propagator for the spin liquid.  At long distances,
this propagator is dominated by the relativistic structure of the
excitations around the Dirac point at ${\bf K}$, where $\gamma_{s\bk
+{\bf K}} = ic (k_{x}+ik_{y})$ ($c = 3K/2$
). The approximate structure
of $\Sigma (x,x')$ can be obtained by power-counting: since ${\cal G} (k)\sim
1/k$ in Fourier space,  ${\cal G} (x)\sim 1/x^{2}$, so we expect that 
\begin{equation}\label{}
\Sigma (x,x')\sim 
\left[
\sigma^{a}{\cal V} (x)
\frac{1}{|x-x'|^{2}} {\cal V}\dg (x')\sigma^{a}  \right]
e^{i {\bf K}\cdot(\bx-\bx')}
\end{equation}
where $|x|^{2}= x^{2}- c^{2}t^{2}$. 
A more detailed calculation gives 
\begin{equation}\label{}
{\cal G} (\vec{ x},t) = \frac{
c t-\vec{ x}\cdot \vec{\beta }}{4 \pi|x|^{3}}e^{i {\bf K}\cdot\bx}
\end{equation}
where $\vec{\beta }=  (\alpha_{y},\alpha_{x})$ defines the sublattice
structure of the tunneling.

This infinite-range,
power-law decay of the tunneling amplitude means that in the
ground-state, the tunneling electrons sample the fractionalized order
at arbitrarily large distances.   In this way, we see that the
development of a decoupled, coherent, neutral Dirac cone is a direct consequence 
of the fractionalization of the order at infinite length scales. 

\section{ { Statistical Mechanics } and Long Wavelength action}\label{sec8}

\subsection{{Statistical Mechanics and } Phase Diagram }\label{pdiax}
We now discuss the statistical mechanics and long-wavelength of the
order fractionalized phase. 
\figwidth=\columnwidth
\fg{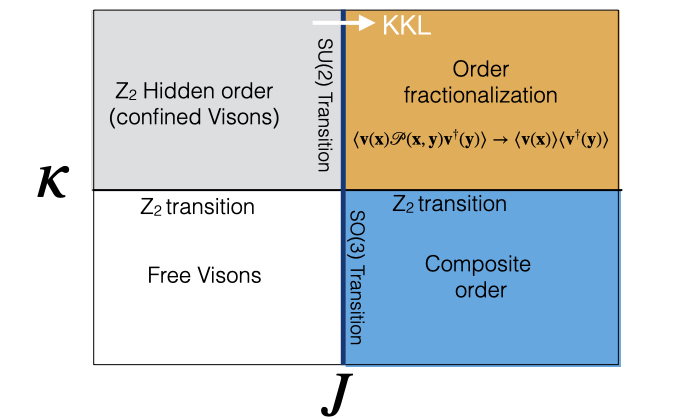}{phasedia2}{Schematic Phase diagram for the
3D classical gauged spinor model from \eqref{action}. {White arrow
shows the order-fractionalization transition.}}
When we integrate out the Fermions from the model, we are left with 
a $Z_{2}$ lattice gauge theory of the spin liquid, 
coupled to the 
 ``matter
fields'' provided by the Kondo spinors $v_{j}$
\begin{equation}\label{action}
H= -J\sum_{(i,j)} v\dg  _{i}u_{ij}v_{j} + U \sum_{j}
(v\dg_{j}v_{j}-1)^{2} 
- \kappa\sum_{p} \prod_{\square}u_{(l,m)}. 
\end{equation}
where the the $U$ term constrains the $v_{j}$  
to fixed magnitude and final plaquette
term ascribes an energy cost of 
$2\kappa$ to each vison.  Here 
we have used our earlier notation $u_{(l,m)}$
where the parentheses orders the site indices so that the $A$ sublattice is first. {The condensed 
$v$-spinors 
are the Higgs fields for the $Z_{2}$ gauge field $u$, transforming its
uncontractible Wilson loops into ``Kondo domain walls'' of finite energy
density. (Fig. \ref{figdom})}
At finite
temperatures, we expect no phase transitions, for the orientational
degrees of freedom are eliminated by the Mermin Wagner theorem, and the
presence of small but finite concentrations of visons eliminates the
possibility of a $Z_{2}$ phase transition \footnote{Note that the 
gauged Ising model theory in two dimensions is the Kramers Wannier 
dual of a conventional Ising model
in a magnetic field that replaces the phase transition by a cross-over
\cite{balian75}.}. 

{There is an interesting issue to what extent charge $e$ bosons may
survive as excitations at nonzero $T$. This depends on the ratio of
$J$ to the vison gap of order $\kappa$. Consider
the average of a large Wilson loop of area $S$. Its value for a
configuration with $n$ visons is $(-1)^n$. At $J_K=0$ the gauge field
is not higgsed, the excitations are point-like visons. The probability
of such a configuration is $\sim C_S^n \exp[-n \kappa/T]$, where
$C_{S}^{n}=n!(S-n)!/S!$ is a combinatoric pre-factor. The sum
over configurations gives 
\begin{equation}
\langle W(S)\rangle  = \sum_{n} (-1)^{n} C^{n}_{S}= (1-
\exp[-\kappa/T])^S = e^{-S/\pi R^{2}}
\end{equation}
giving rise to an area-law confinement at finite T with a confinement
area 
$\pi R^{2}\sim 1/{\ln (1-e^{-\kappa/T})}\sim e^{\kappa/T}$ proportional to the inverse vison density. On the other hand the
correlation  length $\xi$ of the $\vu$-fields  calculated in the limit
$\kappa \rightarrow \infty$ is $\xi \sim \exp(J/T)$. So, if $J \ltappr
\kappa$, then $\xi \ll R$ and 
finite-temperature 
confinement effects are superceded by the finite
orientational correlation length of the order parameter  and are thus
unimportant.  }

However, 
 the quantum model for the Kitaev Kondo lattice at { \sl zero
temperature} is
equivalent to the statistical mechanics of the above 
$Z_{2}$ gauged spinor lattice in 3D \cite{balian75}. 
In the absence of the matter fields, the pure 3D $Z_{2}$ lattice gauge theory 
develops an Ising phase transition for sufficiently large $\kappa$,
into a phase where visons are
absent\cite{fradkinshenker}. 
At small finite $J$ the Ising
transition persists, even though the 
orientational order of the spinors will be absent, 
corresponding to an entry 
into the Higg's phase of the gauged Ising
model.  { This phase corresponds to
the Yao Lee model at zero temperature. 
(Grey region of Fig. \ref{phasedia2}).}
At still larger $J$, the $\hat v$
will develop orientational  order 
in which $\Sigma_{ab} (x,y)\sim {\cal V} (x){\cal V}\dg (y)$
factorizes at long distances, {corresponding to a state of
fractionalized order. This is the phase transition described by our
mean-field theory (Orange region of
Fig. \ref{phasedia2})}. 
Lastly, we note that in the region of large $J$, but small $\kappa$,
{(which is not applicable to the Kitaev Kondo model),}  the local quantity 
$\Sigma_{ab} (x ,x )$ is expected to develop long range order in the 
presence of deconfined visons, corresponding to unfractionalized, 
vector order. { In this phase, the order parameter is the
unfractionalized  composite order parameter \eqref{triad}. (Blue
region of Fig. \ref{phasedia2})}. The conjectured phase diagram for the model is shown in Fig. \ref{phasedia2}. 

\subsection{Long Wavelength action }\label{}

A discussion of the long-wavelength action of the order fractionalized
phase is simplified by taking
the 
special case, where $K=t$, leading to a relativistic field theory 
with an effective speed of light $c_{E} =
(3/2) K$ governing all excitations {(see \eqref{Hfull2})}.  So long as there are no domain
walls, the relativistic, coarse-grained action for 
slow variations in the spinor ${v} (\bx,\tau )$ is
\begin{eqnarray}\label{}
S = {\rho_{s}}\int d^2xd\tau \left|
\left(\partial_{\mu}+i \frac{e}{\hbar }A_{\mu}
\right){\vu}\right|^{2}\\ \nonumber
\end{eqnarray}
where $\partial_{\mu}= (1/c_{E}\partial_{\tau },\vec{\nabla })$. 
It is convenient to rewrite this in terms of the four-component spinor
${\cal Z}$, as
\begin{eqnarray}\label{}
S = {\rho_{s}}\int d^2xd\tau \left|
\left(\partial_{\mu}+i \frac{e}{\hbar }A_{\mu}\tau_{3}
\right){\cal Z}\right|^{2}\\ \nonumber
\end{eqnarray}
If we write ${\cal Z}$ in terms of Euler angles, 
\begin{equation}\label{}
{\cal Z} = 
\exp \left[-i \frac{\phi }{2}\tau_{3} \right]
\exp \left[-i \frac{\theta }{2}\tau_{2} \right]
\exp \left[-i \frac{\psi }{2}\tau_{3} \right]\cal Z_{0},
\end{equation}
then $\partial_{\mu}{\cal Z} = -\frac{i}{2}
({\omega^{a}})_{\mu}\sigma_{a}{\cal Z}$ defines the components of the
angular velocity $\vec{\omega}_{\mu}= {\omega^{a}}_{\mu} {\bf d}^{a}$ 
 measured in the body-axis frame, i.e $\partial_{\mu}{\bf d}^{b} =
 \vec{ \omega}_{\mu}\times {\bf d}^{b} $.  It follows that
\begin{equation}\label{}
\left(i\partial_{\mu} - \frac{e}{\hbar }A_{\mu}\tau_{3} \right){\cal
Z} = \frac{1}{2}(\omega_{\mu}^{a}- \frac{2e}{\hbar }A_{\mu}\delta^{a3}) \tau_{a}{\cal Z},
\end{equation}
allowing us to rewrite the long-wavelength action in the the form of a
principle chiral action, 
\begin{equation}\label{pchiral}
S = \frac{\rho_{s}}{2}\int d^{2}x d\tau \left[(\partial_{\mu}\hat
n)^{2}+ (\omega_{\mu}^{3}- \frac{2e}{\hbar }A_{\mu})^{2} \right]
\end{equation}
where we have made the substitution $(\omega^{1}_{\mu})^{2}+
(\omega^{2}_{\mu})^{2}= (\partial_{\mu}\hat n )^{2}$ and $\hat n
\equiv {\bf d}^{3} ={\cal Z}\dg \vec{\sigma }\tau_{3}{\cal Z}$. 

The action \eqref{pchiral} resembles the action of a
superconductor. However, there are a number of important differences. 
\begin{itemize}

\item  In contrast to a superconductor, 
\begin{eqnarray}
J_{\mu } &=&  - \frac{\delta S}{\delta A^{\mu}}= \frac{2e\rho_{s}}{\hbar } (\omega_{\mu}^{3}-\frac{2e}{\hbar
}A_{\mu})\cr
&=& \frac{2e\rho_{s}}{\hbar } (\partial_{\mu}\psi  -\frac{2e}{\hbar
}A_{\mu} + \cos \theta \partial_{\mu}\phi )
\end{eqnarray}
contains an
additional term $ \cos \theta \partial_{\mu}\phi $, derived from rotations of $\hat n$, 
so the magnetic aspects of the phase associated with $\hat n$ 
are  intertwined with the superconducting properties. 

\item The first homotopy class $\pi_{1} (SU (2))=0$ is empty, implying
that there are no topologically stable vortices of an SU(2) order
parameter.  
Thus in
general, any current loop can be relaxed by 
relaxed by the rotation of the magnetic vector $\hat n$ out of the
plane. 
Magnetic anisotropy is required to stabilize superfuid or
superconducting behavior\cite{coleman94}. 

\item 
Although the vorticity of a screening current has no topological
protection, for the case of charged conduction electrons, 
a fragile Meissner phase is expected\cite{Babaev2014}, because the
relaxation of surface screening currents requires the passage of
skyrmions into the condensate. The energy of a single skyrmion is
$4\pi\rho_s$, so their penetration into the bulk
needs to offset by a finite field external field. Thus we expect that below a critical field,
this paired state will exhibit a fragile Meissner effect\cite{Erten:2017bj}. 

\end{itemize}

Finally we note that a magnetic field introduces a Zeeman coupling
to the electrons and the underlying spin liquid. The Yao-Lee spin
liquid now acquires a Fermi surface.  The effective action now contains 
terms of the form $-g\mu_{B}v\dg \sigma_{z}v$, which
convert the physics into that of an $x-y$ model with a finite temperature BKT
transition associated with the binding of  vortices.
One of the interesting questions, is whether this state will exhibit
$h/e$ vortices characteristic of a charge $e$ condensate?
In fact, the development of these 
vortices 
depends subtlely on the energetics of the visons\cite{sachdev92}. 
The composite order parameter 
\begin{equation}\label{}
\langle v^T (\bx )i\sigma_{2}\vec{ \sigma }v (\bx )\rangle \sim (\hat
d_{1} (\bx)+ i d_{2} (\bx))
\end{equation}
carries charge $2e$.  If we rotate the vectors $d_{1} (x)$ and $\hat
d_{2} (x)$ through $2 \pi$
 about the $\hat d_{3} (x)$ axis, we create an
$h/2e$ vortex. Such a vortex rotates the underlying spinor
order through $2\pi$, causing it to pick up a minus sign, so that two
$h/2e$ vortices are connected by  a
Kondo domain wall whose energy grows with length, which would a priori
bind two $h/2e$ vortices into a single $h/e$ vortex. However, the
naked domain wall  can be removed by binding a vison to the 
the $h/2e$ vortex.  The confinement of the $h/2e$ vortices into 
$h/e$ vortices thus depends on whether the binding energy is negative, or
\begin{equation}\label{}
E\biggl[\frac{h}{e}\biggr]- 2\left(E\biggl[\frac{h}{2e}\biggr] +E_{V}  \right)<0
\end{equation}
For small enough superfluid stiffness, this quantity is necessarily
negative, so that in the vicinity of the quantum phase transition into
the order-fractionalized state, we expect $h/e$ vortices to be stable in
the fractionalized condensate. 

\section{Discussion: Broader Implications}\label{sec9}

We have presented a model realization of order
fractionalization in a Kondo lattice where
conduction electrons interact with a $Z_{2}$ spin liquid. Our theory, which
describes the interaction of an  emergent Z$_2$ gauge theory with matter, 
has several
distinct features:
\begin{enumerate}
\item The order parameter, a spinor, 
carries charge $e$ and  spin S=1/2. 

\item The broken symmetry state has a gapless Majorana
mode in the bulk, which results from a mismatch between the quantum numbers of the
conduction electrons, which carry spin 1/2, charge $e$,  and the elementary spin-one
majorana excitations of the $Z_{2}$ spin liquid. 
This mismatch determines the quantum
numbers of the order parameter formed as a bound-state between conduction
electrons and Majorana fermions. 

\item Fractionalized order, in which the spinor order parameter
develops long-range order,  allows the electrons to
coherently tunnel through the spin liquid over arbitrarily long distances.

\end{enumerate}

The condensation of an order parameter carrying a Z$_2$ charge 
is a direct consequence of the massive $Z_{2}$ gauge field, which eliminates
visons and gives rise to deconfined Majorana fermions in the spin
liquid. Although the models we have discussed,
involve a static Z$_2$ gauge field, whose excitations - visons, are immobile, the phenomena
we observe in our model only requires 
that the underlying spin-liquid contains gapped gauge excitations.

  Some features of our model  are related to its low
  dimensionality. 
In 2D fractionalized order is only strictly present at zero
temperature when there are no visons, however, as we discussed in Section \ref{pdiax}, vestiges of
order fractionalization order will persist to finite temperatures provided the
correlation length $\xi $ of the order parameter is shorter than the
confinement radius $R$ of the gauge field determined by the density of
thermally excited visons.  Order fractionalization is likely to become
more robust in higher dimensions. In particular, like the Kitaev spin
liquid, our model serves as a platform { for an entire family of three
dimensional lattices with trivalent co-ordination\cite{eschmann}, including the 
hyperoctagonal lattice (the subject of a forthcoming paper \cite{forthcoming})} where the 
phase transition occurs at finite temperature and at arbitrary small $J_K$ so that all analytical calculations can be performed in a controllable manner.

More generally, we expect that the order fractionalization observed in
our model constitutes an emergent phenomenon with physically observable
consequences in the quantum universe at large, including quantum materials, and
in the context of relativistic theories (see section \ref{sec5}). 
This wider context also includes the
expectation of superconducting phases with
gapless Majorana Fermi surfaces, and $Z_{2}$ phase transitions in
which the domain walls associated with an emergent spinor order may
manifest themselves as hidden order phase transitions. Several aspects
of quantum materials, including heavy fermion compounds with hidden
order, such as URu$_{2}$Si$_{2}$\cite{Mydosh:2011wm}, and superconductors and insulators
with signs of an underlying Fermi surface, such as UTe$_{2}$\cite{Ran:2019dj} and SmB$_{6}$\cite{Hartstein_2017}
are interesting candidates for these novel possibilities.

%
%


{\it Acknowledgements}

This work was supported by Office of Basic Energy Sciences, Material
 Sciences and Engineering Division, U.S. Department of Energy (DOE)
 under Contracts No. DE-SC0012704 (AMT) and DE-FG02-99ER45790 (PC).
{AMT and PC contributed equally to this work.} 
AMT is grateful to Lukas Janssen for an excellent presentation of their results on the YLK model and to Joerg Schmalian to valuable remarks.
PC acknowledges discussions with Tom Banks, Premi Chandra, Eduardo
Fradkin, Yashar Komijani  and Subir Sachdev, and the
support of the Aspen Center for Physics under NSF Grant PHY-1607611,
where part of this work was completed. 

\appendix

\begin{widetext}
\section{ Alternative Fermionization of the Yao-Lee model using Jordan-Wigner
Fermions }\label{AppendixA}

Like the Kitaev honeycomb model, the Yao-Lee model can be solved using
a Jordan-Wigner transformation. This alternative fermionization scheme
allows a derivation of the model that does not involve an expansion of
the Hilbert space. 
The derivation here is an adaptation of that of Feng, Zhang and
Xiang\cite{feng2007} for the Kitaev honeycomb model, that incorporates
the additional degrees of freedom in Yao-Lee model.
To see how this works, we first redraw the honeycomb
as a brick-wall lattice,
composed of one dimensional chains with alternating cross-links (see
Fig. \ref{fig2app} A.), where the horizontal 
chains are labelled by the index $l$ , and
the
position along the chains is labelled by the index $j$. 
The Yao-Lee Hamiltonian with antiferromagnetic bond-strengths $K^{x}$, $K^{y}$
and $K^{z}$ is written
\begin{equation}\label{defYL}
H =  \frac{1}{2}\sum_{l+j\in \hbox{even}} \biggl[
K^{x}
(\lambda_{jl}^{x}\lambda_{j-1,l}^{x}
)\vec{\sigma }_{jl}\cdot \vec{\sigma }_{j-1,l}
 + 
K^{y}
(\lambda_{jl}^{y}\lambda_{j+1,l}^{y}
)\vec{\sigma }_{jl}\cdot \vec{\sigma }_{j+1,l} + 
K^{z}
(\lambda_{jl}^{z}\lambda_{j,l+1}^{z}
) \vec{\sigma }_{jl}\cdot \vec{\sigma }_{j,l+1}
\biggr],
\end{equation}
where the $\vec{\lambda}_{jl}$ and $\vec{\sigma}_{jl}$ are Pauli
matrices for the orbital, and spin degrees of freedom, respectively.
We label the sites so that $A$ sublattice $j+l$ is odd, while on the
$B$ sublattice, $j+l$ is even. 

\end{widetext}

\figwidth=\columnwidth
\fgh{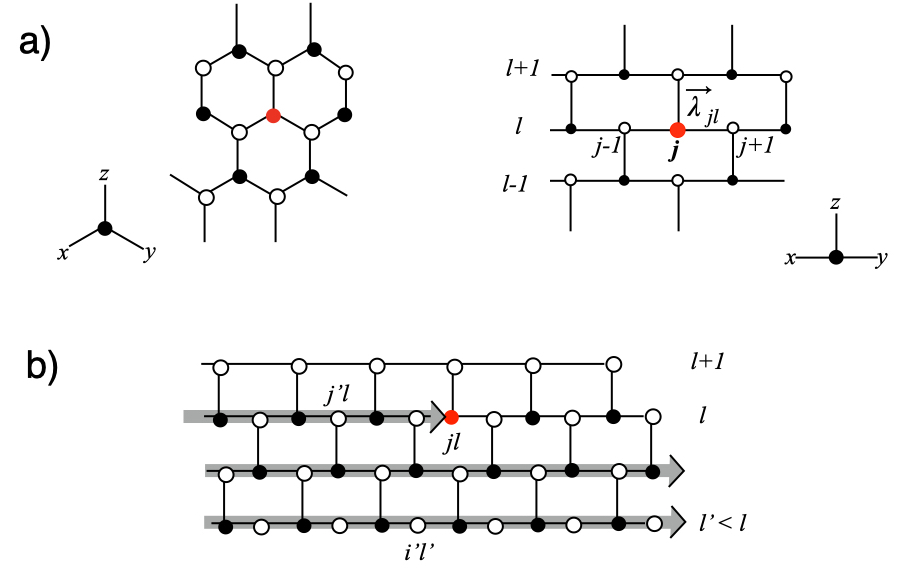}{fig2app}{a) Equivalence between honeycomb and
``brick-wall'' lattice. ``A'' sites (white circles) correspond to odd
$j+l$, whereas ``B'' sites (black circles) correspond to even $j+l$.
b) Showing Jordan-Wigner string ${\cal  P}_{jl}=\exp [i\Phi_{jl}]$
snaking along the rows up to site $(j,l)$.}

Following \cite{feng2007}, we carry out  a Jordan-Wigner
transformation on the frustrated orbital degrees of freedom $\lambda^{a}_{jl}$,  as follows:
\begin{eqnarray}\label{l}
\lambda_{jl}^{+}&=& (\lambda_{jl}^{x}+i \lambda_{jl}^{y})/2 = f\dg_{jl}
{P}_{jl}
\cr
\lambda_{jl}^{-}&=& (\lambda_{jl}^{x}-i \lambda_{jl}^{y})/2 = f_{jl}
{P}_{jl }
\cr
\lambda_{jl}^{z}&=& 2 n_{jl}-1,
\end{eqnarray}
where ${P}_{jl}=\exp [i\Phi_{jl}]$ is a real, 
$Z_{2}$ string
operator, where 
\begin{equation}\label{}
\Phi_{jl} = \pi \sum_{j'<j}n_{j'l}
+\pi \sum_{i}\sum_{l'<l,}
n_{il'}.
\end{equation}
The string operator ${P}_{jl}$
runs from left to right
along each row of the lattice, starting at the bottom left-hand
corner, continuing until it reaches site $(j,l)$
(see Fig. \ref{fig2app}
B. )
${ P}_{jl}$ 
commutes with the fermions $f_{mn}$ at sites above or to
the right of site $(j,l)$, but {\sl anticommutes} with all fermions 
$f_{mn}$ along its path, i.e sites to the left of $jl$ on the same
row, and sites on any row below row $l$. This 
guarantees the the orbital operators $\lambda^{a}_{jl}$ commute
between sites. 
Applying the Jordan-Wigner transformation to the orbital interaction
terms in \eqref{defYL}, we have 
\begin{eqnarray}\label{}
\lambda^{x}_{jl} \lambda^{x}_{j-1,l}
&=& (f_{j,l}+ f\dg_{j,l})(f_{j-1,l}- f\dg_{j-1,l}),\cr
\lambda^{y}_{j+1,l} \lambda^{y}_{j,l}
&=& (f\dg _{j+1,l}- f_{j+1,l})(f\dg _{j,l}+ f_{j,l}), \cr
\lambda^{z}_{jl}\lambda^{z}_{j,l+1} &=& (2 n_{jl}-1) (2n_{j,l+1}-1),
\end{eqnarray}
so that the fermionized Hamiltonian becomes
\begin{eqnarray}\label{l}
H &=&  \frac{1}{2}\sum_{l+j\in \hbox{even}} \biggl[
K^{x} (f_{jl}+f_{jl}\dg ) (f_{j-1l}- f\dg _{j-1l})
(\vec{\sigma }_{jl}\cdot\vec{\sigma}_{j-1l})\cr
&+& 
K^{y}
(f\dg_{j+1l}-f_{j+1l})
(f\dg_{jl}+f_{jl})(\vec{\sigma }_{jl}\cdot\vec{\sigma}_{j+1l})
\cr
& +& 
K^{z}
(2n_{jl}-1) (2n_{j,l+1}-1)(\vec{\sigma }_{jl}\cdot\vec{\sigma}_{jl+1})
\biggr].
\end{eqnarray}
Splitting the fermions into their Majorana components, choosing 
$f_{jl}= (c_{jl}\red{-}ib_{jl})/\sqrt{2}$ for even $j+l$, while
$f_{jl}= (b_{jl}+ic_{jl})/\sqrt{2}$ for odd $j+1$, so that 
$c_{jl}= ( f_{jl}+f\dg_{jl})/\sqrt{2}$ (even $j+l$) and $
c_{j,l}
= i( f_{j,l}-f\dg_{j,l})/ \sqrt{2}$ (odd $j+l$) and 
\begin{eqnarray}\label{l}
(2n_{jl}-1) (2n_{j,l+1}^{z}-1)= 
ic_{j,l+1}u_{jl+1,jl}
c_{jl},
\end{eqnarray}
where  $u_{jl+1,jl}= \red{-}2i 
b_{jl+1}b_{jl}$ is a $Z_{2}$ field operator with eigenvalues $\pm 1$ that lives
on the vertical $z$ bonds.   The Hamiltonian can then be written
\begin{eqnarray}\label{jordwig}
H &=& \sum_{j+l\in \hbox{even}}
i (K^{x}  c_{j-1l}\vec{\sigma }_{j-1l} + K^{y}c_{j+1l}\vec{\sigma
}_{j+1l}\cr
 &+& K^{z}c_{jl+1}\vec{\sigma }_{jl+1}u_{jl+1,jl}
)\cdot c_{jl}\vec{\sigma }_{jl}.
\end{eqnarray}
Notice that the operators $u_{jl+1,jl}=-2ib_{j,l+1}b_{j,l}$ only appear on
vertical bonds in the Hamiltonian, commuting 
with the entire Hamiltonian, forming static $Z_{2}$
gauge fields. 

Finally, we note that the operators 
\begin{eqnarray}\label{l}
\vec{\chi }_{jl}= c_{jl}\vec{\sigma}_{jl}
\end{eqnarray}
are real, and satisfy canonical
anticommutation relations
\begin{equation}\label{}
\chi^{a}_{jl}= (\chi^{a}_{jl})\dg, \qquad
\{\chi^{a}_{jl},\chi^{b}_{mn} \} = \delta^{ab}\delta_{jm}\delta_{ln},
\end{equation}
enabling us to identify them as independent Majorana fermions
(normalized so that $(\chi^{a}_{jl})^{2}=1/2$). 
The Hamiltonian thus reverts to the fermionized version of the Yao-Lee model
\begin{eqnarray}\label{l}
H =\sum_{\langle i,j\rangle } iK^{\alpha_{ij}} u_{i,j}(\vec{\chi
}_{i}\cdot\vec{\chi }_{j}) 
\end{eqnarray}
in the gauge where the gauge fields $u^{x}_{(i,j)}= u^{y}_{(i,j)} =1$
are set to unity in the $x$ and $y$ directions. 
With this  gauge choice
the flux through each hexagon is determined by the vertical bonds
alone. (In our treatment of the model, we set all $K^{a}=K$ to be equal.)
\fg{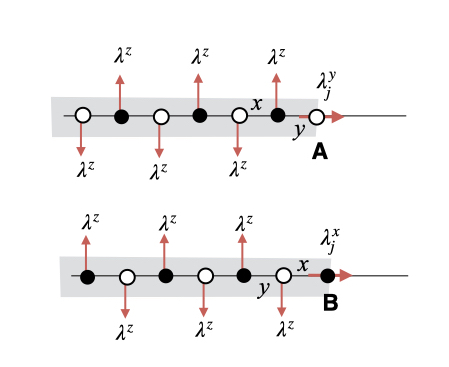}{fig3app}{
The modified string operator ${\cal P}_{j}$
incorporates the final link on the string as an extremal bond. }

Lastly, note that we can invert the Jordan-Wigner transformation,
identifying
\begin{equation}\label{}
{P}_{jl}= \prod_{j'l'\in P}  \lambda^{z}_{j'l'}
\end{equation}
as the product of the orbital matrices along the path $P$ of the
string (not including site $(j,l)$
). The $\chi$ majoranas can then be
written as
\begin{equation}\label{express2}
\frac{1}{\sqrt{2}}\vec{\chi }_{jl} = \vec{S}_{jl}
{P}_{jl}
\times \left\{ \begin{array}{cc}
\lambda^{y}_{jl}, &(\hbox{A site}\ j+l  \in \hbox{odd}),\cr
\lambda^{x}_{jl}, &(\hbox{B site}\  j+l  \in \hbox{even}),
\end{array}\right. 
\end{equation}
We can incorporate the dangling $\lambda^{a}$ operators by regarding the
final link on the string as an extremal bond, defining
\begin{equation}\label{}
{\cal P}_{jl} = \lambda_{jl}^{\alpha_{jl,j-1l}}P_{jl}
\end{equation}
so that now
\begin{equation}\label{}
\frac{1}{\sqrt{2}}\vec{\chi }_{jl} = \vec{S}_{jl}
{\cal P}_{jl} 
\end{equation}
providing a unique, non-local expression for the Majorana spin
excitations.

%

 \end{document}